%-------------------------------------------------------------------
% MNSAMPLE.TEX
%
% A sample plain TeX single/two column Monthly Notices article.
%
% v1.5  --- released 25th August 1994 (M. Reed)
% v1.4  --- released 22nd February 1994
% v1.3  --- released  8th December 1992
%
% Copyright Cambridge University Press

% The following line automatically loads the mn macros if you are not
% using a format file.
\ifx\mnmacrosloaded\undefined % MN.TEX (Computer Modern version)
%
% plain TeX single / double column macros for the
% Monthly Notices of Royal Astronomical Society
%
% v1.5  (mn.tex)  --- released 25th August 1994 (M. Reed)
% v1.4      "     --- released 22nd February 1994
% v1.3  (mnd.tex) --- released 28th November 1992
% v1.26     "     --- released  1st August 1992
% v1.25     "     --- released 25th February 1992
%
% Copyright Cambridge University Press
%
% > Incorporating special symbol code from laa.sty v1.1 (25th Feb 1991)
%   used with the permission of Springer Verlag.
% > Incorporating parts of mssymb.tex (8th July 1987).
% > Incorporating NewFont.sty v ALPHA patchlevel 8 (16th August 1994).

\catcode `\@=11 % @ signs are letters

\def\@version{1.5}
\def\@verdate{25th Aug 1994}

% Fonts: Computer Modern / Monotype Times (CUP only)
%
% Font family sizes available:
%   8pt, 9pt, 10pt, 11pt, 14pt and 17pt.
%
% Faces available:
%   \rm, math italic, symbol, \it, \bf, \sl, \tt, \sc, \sf, \cal, \em,
%   \mit and \oldstyle.

% define the typeface in use

\newif\ifprod@font

\ifx\@typeface\undefined
  \def\@typeface{Comp. Modern}\prod@fontfalse
\else
  \prod@fonttrue % We want Times
\fi

\def\newfam{\alloc@8\fam\chardef\sixt@@n} % made not outer

\ifprod@font
\font\fiverm=mtr10 at 5pt
\font\fivebf=mtbx10 at 5pt
\font\fiveit=mtti10 at 5pt
\font\fivesl=mtsl10 at 5pt
\font\fivett=cmtt8 at 5pt     \hyphenchar\fivett=-1
\font\fivecsc=mtcsc10 at 5pt
\font\fivesf=mtss10 at 5pt
\font\fivei=mtmi10 at 5pt      \skewchar\fivei='177
\font\fivesy=mtsy10 at 5pt     \skewchar\fivesy='60

\font\sixrm=mtr10 at 6pt
\font\sixbf=mtbx10 at 6pt
\font\sixit=mtti10 at 6pt
\font\sixsl=mtsl10 at 6pt
\font\sixtt=cmtt8 at 6pt      \hyphenchar\sixtt=-1
\font\sixcsc=mtcsc10 at 6pt
\font\sixsf=mtss10 at 6pt
\font\sixi=mtmi10 at 6pt       \skewchar\sixi='177
\font\sixsy=mtsy10 at 6pt      \skewchar\sixsy='60

\font\sevenrm=mtr10 at 7pt
\font\sevenbf=mtbx10 at 7pt
\font\sevenit=mtti10 at 7pt
\font\sevensl=mtsl10 at 7pt
\font\seventt=cmtt8 at 7pt     \hyphenchar\seventt=-1
\font\sevencsc=mtcsc10 at 7pt
\font\sevensf=mtss10 at 7pt
\font\seveni=mtmi10 at 7pt      \skewchar\seveni='177
\font\sevensy=mtsy10 at 7pt     \skewchar\sevensy='60

\font\eightrm=mtr10 at 8pt
\font\eightbf=mtbx10 at 8pt
\font\eightit=mtti10 at 8pt
\font\eighti=mtmi10 at 8pt      \skewchar\eighti='177
\font\eightsy=mtsy10 at 8pt     \skewchar\eightsy='60
\font\eightsl=mtsl10 at 8pt
\font\eighttt=cmtt8             \hyphenchar\eighttt=-1
\font\eightcsc=mtcsc10 at 8pt
\font\eightsf=mtss10 at 8pt

\font\ninerm=mtr10 at 9pt
\font\ninebf=mtbx10 at 9pt
\font\nineit=mtti10 at 9pt
\font\ninei=mtmi10 at 9pt      \skewchar\ninei='177
\font\ninesy=mtsy10 at 9pt     \skewchar\ninesy='60
\font\ninesl=mtsl10 at 9pt
\font\ninett=cmtt9             \hyphenchar\ninett=-1
\font\ninecsc=mtcsc10 at 9pt
\font\ninesf=mtss10 at 9pt

\font\tenrm=mtr10
\font\tenbf=mtbx10
\font\tenit=mtti10
\font\teni=mtmi10		\skewchar\teni='177
\font\tensy=mtsy10		\skewchar\tensy='60
\font\tenex=cmex10
\font\tensl=mtsl10
\font\tentt=cmtt10		\hyphenchar\tentt=-1
\font\tencsc=mtcsc10
\font\tensf=mtss10

\font\elevenrm=mtr10 at 11pt
\font\elevenbf=mtbx10 at 11pt
\font\elevenit=mtti10 at 11pt
\font\eleveni=mtmi10 at 11pt      \skewchar\eleveni='177
\font\elevensy=mtsy10 at 11pt     \skewchar\elevensy='60
\font\elevensl=mtsl10 at 11pt
\font\eleventt=cmtt10 at 11pt     \hyphenchar\eleventt=-1
\font\elevencsc=mtcsc10 at 11pt
\font\elevensf=mtss10 at 11pt

\font\twelverm=mtr10 at 12pt
\font\twelvebf=mtbx10 at 12pt
\font\twelveit=mtti10 at 12pt
\font\twelvesl=mtsl10 at 12pt
\font\twelvett=cmtt12             \hyphenchar\twelvett=-1
\font\twelvecsc=mtcsc10 at 12pt
\font\twelvesf=mtss10 at 12pt
\font\twelvei=mtmi10 at 12pt      \skewchar\twelvei='177
\font\twelvesy=mtsy10 at 12pt     \skewchar\twelvesy='60

\font\fourteenrm=mtr10 at 14pt
\font\fourteenbf=mtbx10 at 14pt
\font\fourteenit=mtti10 at 14pt
\font\fourteeni=mtmi10 at 14pt      \skewchar\fourteeni='177
\font\fourteensy=mtsy10 at 14pt     \skewchar\fourteensy='60
\font\fourteensl=mtsl10 at 14pt
\font\fourteentt=cmtt12 at 14pt     \hyphenchar\fourteentt=-1
\font\fourteencsc=mtcsc10 at 14pt
\font\fourteensf=mtss10 at 14pt

\font\seventeenrm=mtr10 at 17pt
\font\seventeenbf=mtbx10 at 17pt
\font\seventeenit=mtti10 at 17pt
\font\seventeeni=mtmi10 at 17pt      \skewchar\seventeeni='177
\font\seventeensy=mtsy10 at 17pt     \skewchar\seventeensy='60
\font\seventeensl=mtsl10 at 17pt
\font\seventeentt=cmtt12 at 17pt     \hyphenchar\seventeentt=-1
\font\seventeencsc=mtcsc10 at 17pt
\font\seventeensf=mtss10 at 17pt
\else
\font\fiverm=cmr5
\font\fivei=cmmi5             \skewchar\fivei='177
\font\fivesy=cmsy5            \skewchar\fivesy='60
\font\fivebf=cmbx5

\font\sixrm=cmr6
\font\sixi=cmmi6             \skewchar\sixi='177
\font\sixsy=cmsy6            \skewchar\sixsy='60
\font\sixbf=cmbx6

\font\sevenrm=cmr7
\font\sevenit=cmti7
\font\seveni=cmmi7             \skewchar\seveni='177
\font\sevensy=cmsy7            \skewchar\sevensy='60
\font\sevenbf=cmbx7

\font\eightrm=cmr8
\font\eightbf=cmbx8
\font\eightit=cmti8
\font\eighti=cmmi8			\skewchar\eighti='177
\font\eightsy=cmsy8			\skewchar\eightsy='60
\font\eightsl=cmsl8
\font\eighttt=cmtt8			\hyphenchar\eighttt=-1
\font\eightcsc=cmcsc10 at 8pt
\font\eightsf=cmss8

\font\ninerm=cmr9
\font\ninebf=cmbx9
\font\nineit=cmti9
\font\ninei=cmmi9			\skewchar\ninei='177
\font\ninesy=cmsy9			\skewchar\ninesy='60
\font\ninesl=cmsl9
\font\ninett=cmtt9			\hyphenchar\ninett=-1
\font\ninecsc=cmcsc10 at 9pt
\font\ninesf=cmss9

\font\tenrm=cmr10
\font\tenbf=cmbx10
\font\tenit=cmti10
\font\teni=cmmi10		\skewchar\teni='177
\font\tensy=cmsy10		\skewchar\tensy='60
\font\tenex=cmex10
\font\tensl=cmsl10
\font\tentt=cmtt10		\hyphenchar\tentt=-1
\font\tencsc=cmcsc10
\font\tensf=cmss10

\font\elevenrm=cmr10 scaled \magstephalf
\font\elevenbf=cmbx10 scaled \magstephalf
\font\elevenit=cmti10 scaled \magstephalf
\font\eleveni=cmmi10 scaled \magstephalf	\skewchar\eleveni='177
\font\elevensy=cmsy10 scaled \magstephalf	\skewchar\elevensy='60
\font\elevensl=cmsl10 scaled \magstephalf
\font\eleventt=cmtt10 scaled \magstephalf	\hyphenchar\eleventt=-1
\font\elevencsc=cmcsc10 scaled \magstephalf
\font\elevensf=cmss10 scaled \magstephalf

\font\twelverm=cmr10 scaled \magstep1
\font\twelvebf=cmbx10 scaled \magstep1
\font\twelvei=cmmi10 scaled \magstep1      \skewchar\twelvei='177
\font\twelvesy=cmsy10 scaled \magstep1     \skewchar\twelvesy='60

\font\fourteenrm=cmr10 scaled \magstep2
\font\fourteenbf=cmbx10 scaled \magstep2
\font\fourteenit=cmti10 scaled \magstep2
\font\fourteeni=cmmi10 scaled \magstep2		\skewchar\fourteeni='177
\font\fourteensy=cmsy10 scaled \magstep2	\skewchar\fourteensy='60
\font\fourteensl=cmsl10 scaled \magstep2
\font\fourteentt=cmtt10 scaled \magstep2	\hyphenchar\fourteentt=-1
\font\fourteencsc=cmcsc10 scaled \magstep2
\font\fourteensf=cmss10 scaled \magstep2

\font\seventeenrm=cmr10 scaled \magstep3
\font\seventeenbf=cmbx10 scaled \magstep3
\font\seventeenit=cmti10 scaled \magstep3
\font\seventeeni=cmmi10 scaled \magstep3	\skewchar\seventeeni='177
\font\seventeensy=cmsy10 scaled \magstep3	\skewchar\seventeensy='60
\font\seventeensl=cmsl10 scaled \magstep3
\font\seventeentt=cmtt10 scaled \magstep3	\hyphenchar\seventeentt=-1
\font\seventeencsc=cmcsc10 scaled \magstep3
\font\seventeensf=cmss10 scaled \magstep3
\fi

\def\hexnumber#1{\ifcase#1 0\or1\or2\or3\or4\or5\or6\or7\or8\or9\or
  A\or B\or C\or D\or E\or F\fi}

\def\makestrut{%
  \setbox\strutbox=\hbox{%
    \vrule height.7\baselineskip depth.3\baselineskip width \z@}%
}

\def\baselinestretch{1}
\newskip\tmp@bls

\def\b@ls#1{% set baseline using \baselinestretch as a scale factor
  \tmp@bls=#1\relax
  \baselineskip=#1\relax\makestrut
  \normalbaselineskip=\baselinestretch\tmp@bls
  \normalbaselines
}

\def\nostb@ls#1{% set baseline skip ignoring \baselinestretch
  \normalbaselineskip=#1\relax
  \normalbaselines
  \makestrut
}

% families \itfam, \slfam, \bffam, \ttfam defined in PLAIN.
%
% \itfam is \fam4
% \slfam is \fam5
% \bffam is \fam6
% \ttfam is \fam7

\newfam\scfam  % \fam8
\newfam\sffam  % \fam9

\def\mit{\fam\@ne}
\def\cal{\fam\tw@}
\def\em{\ifdim\fontdimen1\font>\z@ \rm\else\it\fi}

\textfont3=\tenex
\scriptfont3=\tenex
\scriptscriptfont3=\tenex

\setbox0=\hbox{\tenex B} \p@renwd=\wd0 % width of the big left (

\def\eightpoint{% 8^6^5 on 10pt
  \def\rm{\fam0\eightrm}%
  \textfont0=\eightrm \scriptfont0=\sixrm \scriptscriptfont0=\fiverm%
  \textfont1=\eighti  \scriptfont1=\sixi  \scriptscriptfont1=\fivei%
  \textfont2=\eightsy \scriptfont2=\sixsy \scriptscriptfont2=\fivesy%
  \textfont\itfam=\eightit\def\it{\fam\itfam\eightit}%
  \ifprod@font
    \scriptfont\itfam=\sixit
      \scriptscriptfont\itfam=\fiveit
  \else
    \scriptfont\itfam=\eightit
      \scriptscriptfont\itfam=\eightit
  \fi
  \textfont\bffam=\eightbf%
    \scriptfont\bffam=\sixbf%
      \scriptscriptfont\bffam=\fivebf%
  \def\bf{\fam\bffam\eightbf}%
  \textfont\slfam=\eightsl\def\sl{\fam\slfam\eightsl}%
  \ifprod@font
    \scriptfont\slfam=\sixsl
      \scriptscriptfont\slfam=\fivesl
  \else
    \scriptfont\slfam=\eightsl
      \scriptscriptfont\slfam=\eightsl
  \fi
  \textfont\ttfam=\eighttt\def\tt{\fam\ttfam\eighttt}%
  \ifprod@font
    \scriptfont\ttfam=\sixtt
      \scriptscriptfont\ttfam=\fivett
  \else
    \scriptfont\ttfam=\eighttt
      \scriptscriptfont\ttfam=\eighttt
  \fi
  \textfont\scfam=\eightcsc\def\sc{\fam\scfam\eightcsc}%
  \ifprod@font
    \scriptfont\scfam=\sixcsc
      \scriptscriptfont\scfam=\fivecsc
  \else
    \scriptfont\scfam=\eightcsc
      \scriptscriptfont\scfam=\eightcsc
  \fi
  \textfont\sffam=\eightsf\def\sf{\fam\sffam\eightsf}%
  \ifprod@font
    \scriptfont\sffam=\sixsf
      \scriptscriptfont\sffam=\fivesf
  \else
    \scriptfont\sffam=\eightsf
      \scriptscriptfont\sffam=\eightsf
  \fi
  \def\oldstyle{\fam\@ne\eighti}%
  \b@ls{10pt}\rm\@viiipt%
}
\def\@viiipt{}

\def\ninepoint{% 9^6^5 on 11pt (two col) / 12 (single col)
  \def\rm{\fam0\ninerm}%
  \textfont0=\ninerm \scriptfont0=\sixrm \scriptscriptfont0=\fiverm%
  \textfont1=\ninei  \scriptfont1=\sixi  \scriptscriptfont1=\fivei%
  \textfont2=\ninesy \scriptfont2=\sixsy \scriptscriptfont2=\fivesy%
  \textfont\itfam=\nineit\def\it{\fam\itfam\nineit}%
  \ifprod@font
    \scriptfont\itfam=\sixit
      \scriptscriptfont\itfam=\fiveit
  \else
    \scriptfont\itfam=\nineit
      \scriptscriptfont\itfam=\nineit
  \fi
  \textfont\bffam=\ninebf%
    \scriptfont\bffam=\sixbf%
      \scriptscriptfont\bffam=\fivebf%
  \def\bf{\fam\bffam\ninebf}%
  \textfont\slfam=\ninesl\def\sl{\fam\slfam\ninesl}%
  \ifprod@font
    \scriptfont\slfam=\sixsl
      \scriptscriptfont\slfam=\fivesl
  \else
    \scriptfont\slfam=\ninesl
      \scriptscriptfont\slfam=\ninesl
  \fi
  \textfont\ttfam=\ninett\def\tt{\fam\ttfam\ninett}%
  \ifprod@font
    \scriptfont\ttfam=\sixtt
      \scriptscriptfont\ttfam=\fivett
  \else
    \scriptfont\ttfam=\ninett
      \scriptscriptfont\ttfam=\ninett
  \fi
  \textfont\scfam=\ninecsc\def\sc{\fam\scfam\ninecsc}%
  \ifprod@font
    \scriptfont\scfam=\sixcsc
      \scriptscriptfont\scfam=\fivecsc
  \else
    \scriptfont\scfam=\ninecsc
      \scriptscriptfont\scfam=\ninecsc
  \fi
  \textfont\sffam=\ninesf\def\sf{\fam\sffam\ninesf}%
  \ifprod@font
    \scriptfont\sffam=\sixsf
      \scriptscriptfont\sffam=\fivesf
  \else
    \scriptfont\sffam=\ninesf
      \scriptscriptfont\sffam=\ninesf
  \fi
  \def\oldstyle{\fam\@ne\ninei}%
  \b@ls{\TextLeading plus \Feathering}\rm\@ixpt%
}
\def\@ixpt{}

\def\tenpoint{% 10^7^5 on 11pt
  \def\rm{\fam0\tenrm}%
  \textfont0=\tenrm \scriptfont0=\sevenrm \scriptscriptfont0=\fiverm%
  \textfont1=\teni  \scriptfont1=\seveni  \scriptscriptfont1=\fivei%
  \textfont2=\tensy \scriptfont2=\sevensy \scriptscriptfont2=\fivesy%
  \textfont\itfam=\tenit\def\it{\fam\itfam\tenit}%
  \ifprod@font
    \scriptfont\itfam=\sevenit
      \scriptscriptfont\itfam=\fiveit
  \else
    \scriptfont\itfam=\tenit
      \scriptscriptfont\itfam=\tenit
  \fi
  \textfont\bffam=\tenbf%
    \scriptfont\bffam=\sevenbf%
      \scriptscriptfont\bffam=\fivebf%
  \def\bf{\fam\bffam\tenbf}%
  \textfont\slfam=\tensl\def\sl{\fam\slfam\tensl}%
  \ifprod@font
    \scriptfont\slfam=\sevensl
      \scriptscriptfont\slfam=\fivesl
  \else
    \scriptfont\slfam=\tensl
      \scriptscriptfont\slfam=\tensl
  \fi
  \textfont\ttfam=\tentt\def\tt{\fam\ttfam\tentt}%
  \ifprod@font
    \scriptfont\ttfam=\seventt
      \scriptscriptfont\ttfam=\fivett
  \else
    \scriptfont\ttfam=\tentt
      \scriptscriptfont\ttfam=\tentt
  \fi
  \textfont\scfam=\tencsc\def\sc{\fam\scfam\tencsc}%
  \ifprod@font
    \scriptfont\scfam=\sevencsc
      \scriptscriptfont\scfam=\fivecsc
  \else
    \scriptfont\scfam=\tencsc
      \scriptscriptfont\scfam=\tencsc
  \fi
  \textfont\sffam=\tensf\def\sf{\fam\sffam\tensf}%
  \ifprod@font
    \scriptfont\sffam=\sevensf
      \scriptscriptfont\sffam=\fivesf
  \else
    \scriptfont\sffam=\tensf
      \scriptscriptfont\sffam=\tensf
  \fi
  \def\oldstyle{\fam\@ne\teni}%
  \b@ls{11pt}\rm\@xpt%
}
\def\@xpt{}

\def\elevenpoint{% 11^8^6 on 13pt
  \def\rm{\fam0\elevenrm}%
  \textfont0=\elevenrm \scriptfont0=\eightrm \scriptscriptfont0=\sixrm%
  \textfont1=\eleveni  \scriptfont1=\eighti  \scriptscriptfont1=\sixi%
  \textfont2=\elevensy \scriptfont2=\eightsy \scriptscriptfont2=\sixsy%
  \textfont\itfam=\elevenit\def\it{\fam\itfam\elevenit}%
  \ifprod@font
    \scriptfont\itfam=\eightit
      \scriptscriptfont\itfam=\sixit
  \else
    \scriptfont\itfam=\elevenit
      \scriptscriptfont\itfam=\elevenit
  \fi
  \textfont\bffam=\elevenbf%
    \scriptfont\bffam=\eightbf%
      \scriptscriptfont\bffam=\sixbf%
  \def\bf{\fam\bffam\elevenbf}%
  \textfont\slfam=\elevensl\def\sl{\fam\slfam\elevensl}%
  \ifprod@font
    \scriptfont\slfam=\eightsl
      \scriptscriptfont\slfam=\sixsl
  \else
    \scriptfont\slfam=\elevensl
      \scriptscriptfont\slfam=\elevensl
  \fi
  \textfont\ttfam=\eleventt\def\tt{\fam\ttfam\eleventt}%
  \ifprod@font
    \scriptfont\ttfam=\eighttt
      \scriptscriptfont\ttfam=\sixtt
  \else
    \scriptfont\ttfam=\eleventt
      \scriptscriptfont\ttfam=\eleventt
  \fi
  \textfont\scfam=\elevencsc\def\sc{\fam\scfam\elevencsc}%
  \ifprod@font
    \scriptfont\scfam=\eightcsc
      \scriptscriptfont\scfam=\sixcsc
  \else
    \scriptfont\scfam=\elevencsc
      \scriptscriptfont\scfam=\elevencsc
  \fi
  \textfont\sffam=\elevensf\def\sf{\fam\sffam\elevensf}%
  \ifprod@font
    \scriptfont\sffam=\eightsf
      \scriptscriptfont\sffam=\sixsf
  \else
    \scriptfont\sffam=\elevensf
      \scriptscriptfont\sffam=\elevensf
  \fi
  \def\oldstyle{\fam\@ne\eleveni}%
  \b@ls{13pt}\rm\@xipt%
}
\def\@xipt{}

\def\fourteenpoint{% 14^10^7 on 17pt
  \def\rm{\fam0\fourteenrm}%
  \textfont0\fourteenrm  \scriptfont0\tenrm  \scriptscriptfont0\sevenrm%
  \textfont1\fourteeni   \scriptfont1\teni   \scriptscriptfont1\seveni%
  \textfont2\fourteensy  \scriptfont2\tensy  \scriptscriptfont2\sevensy%
  \textfont\itfam=\fourteenit\def\it{\fam\itfam\fourteenit}%
  \ifprod@font
    \scriptfont\itfam=\tenit
      \scriptscriptfont\itfam=\sevenit
  \else
    \scriptfont\itfam=\fourteenit
      \scriptscriptfont\itfam=\fourteenit
  \fi
  \textfont\bffam=\fourteenbf%
    \scriptfont\bffam=\tenbf%
      \scriptscriptfont\bffam=\sevenbf%
  \def\bf{\fam\bffam\fourteenbf}%
  \textfont\slfam=\fourteensl\def\sl{\fam\slfam\fourteensl}%
  \ifprod@font
    \scriptfont\slfam=\tensl
      \scriptscriptfont\slfam=\sevensl
  \else
    \scriptfont\slfam=\fourteensl
      \scriptscriptfont\slfam=\fourteensl
  \fi
  \textfont\ttfam=\fourteentt\def\tt{\fam\ttfam\fourteentt}%
  \ifprod@font
    \scriptfont\ttfam=\tentt
      \scriptscriptfont\ttfam=\seventt
  \else
    \scriptfont\ttfam=\fourteentt
      \scriptscriptfont\ttfam=\fourteentt
  \fi
  \textfont\scfam=\fourteencsc\def\sc{\fam\scfam\fourteencsc}%
  \ifprod@font
    \scriptfont\scfam=\tencsc
      \scriptscriptfont\scfam=\sevencsc
  \else
    \scriptfont\scfam=\fourteencsc
      \scriptscriptfont\scfam=\fourteencsc
  \fi
  \textfont\sffam=\fourteensf\def\sf{\fam\sffam\fourteensf}%
  \ifprod@font
    \scriptfont\sffam=\tensf
      \scriptscriptfont\sffam=\sevensf
  \else
    \scriptfont\sffam=\fourteensf
      \scriptscriptfont\sffam=\fourteensf
  \fi
  \def\oldstyle{\fam\@ne\fourteeni}%
  \b@ls{17pt}\rm\@xivpt%
}
\def\@xivpt{}

\def\seventeenpoint{% 17^12^10 on 20pt
  \def\rm{\fam0\seventeenrm}%
  \textfont0\seventeenrm  \scriptfont0\twelverm  \scriptscriptfont0\tenrm%
  \textfont1\seventeeni   \scriptfont1\twelvei   \scriptscriptfont1\teni%
  \textfont2\seventeensy  \scriptfont2\twelvesy  \scriptscriptfont2\tensy%
  \textfont\itfam=\seventeenit\def\it{\fam\itfam\seventeenit}%
  \ifprod@font
    \scriptfont\itfam=\twelveit
      \scriptscriptfont\itfam=\tenit
  \else
    \scriptfont\itfam=\seventeenit
      \scriptscriptfont\itfam=\seventeenit
  \fi
  \textfont\bffam=\seventeenbf%
    \scriptfont\bffam=\twelvebf%
      \scriptscriptfont\bffam=\tenbf%
  \def\bf{\fam\bffam\seventeenbf}%
  \textfont\slfam=\seventeensl\def\sl{\fam\slfam\seventeensl}%
  \ifprod@font
    \scriptfont\slfam=\twelvesl
      \scriptscriptfont\slfam=\tensl
  \else
    \scriptfont\slfam=\seventeensl
      \scriptscriptfont\slfam=\seventeensl
  \fi
  \textfont\ttfam=\seventeentt\def\tt{\fam\ttfam\seventeentt}%
  \ifprod@font
    \scriptfont\ttfam=\twelvett
      \scriptscriptfont\ttfam=\tentt
  \else
    \scriptfont\ttfam=\seventeentt
      \scriptscriptfont\ttfam=\seventeentt
  \fi
  \textfont\scfam=\seventeencsc\def\sc{\fam\scfam\seventeencsc}%
  \ifprod@font
    \scriptfont\scfam=\twelvecsc
      \scriptscriptfont\scfam=\tencsc
  \else
    \scriptfont\scfam=\seventeencsc
      \scriptscriptfont\scfam=\seventeencsc
  \fi
  \textfont\sffam=\seventeensf\def\sf{\fam\sffam\seventeensf}%
  \ifprod@font
    \scriptfont\sffam=\twelvesf
      \scriptscriptfont\sffam=\tensf
  \else
    \scriptfont\sffam=\seventeensf
      \scriptscriptfont\sffam=\seventeensf
  \fi
  \def\oldstyle{\fam\@ne\seventeeni}%
  \b@ls{20pt}\rm\@xviipt%
}
\def\@xviipt{}

\lineskip=1pt      \normallineskip=\lineskip
\lineskiplimit=\z@ \normallineskiplimit=\lineskiplimit

% BOLD MATH SYMBOLS

\def\loadboldmathnames{%
  \def\balpha{{\bmath{\alpha}}}%
  \def\bbeta{{\bmath{\beta}}}%
  \def\bgamma{{\bmath{\gamma}}}%
  \def\bdelta{{\bmath{\delta}}}%
  \def\bepsilon{{\bmath{\epsilon}}}%
  \def\bzeta{{\bmath{\zeta}}}%
  \def\boldeta{{\bmath{\eta}}}%
  \def\btheta{{\bmath{\theta}}}%
  \def\biota{{\bmath{\iota}}}%
  \def\bkappa{{\bmath{\kappa}}}%
  \def\blambda{{\bmath{\lambda}}}%
  \def\bmu{{\bmath{\mu}}}%
  \def\bnu{{\bmath{\nu}}}%
  \def\bxi{{\bmath{\xi}}}%
  \def\bpi{{\bmath{\pi}}}%
  \def\brho{{\bmath{\rho}}}%
  \def\bsigma{{\bmath{\sigma}}}%
  \def\btau{{\bmath{\tau}}}%
  \def\bupsilon{{\bmath{\upsilon}}}%
  \def\bphi{{\bmath{\phi}}}%
  \def\bchi{{\bmath{\chi}}}%
  \def\bpsi{{\bmath{\psi}}}%
  \def\bomega{{\bmath{\omega}}}%
  \def\bvarepsilon{{\bmath{\varepsilon}}}%
  \def\bvartheta{{\bmath{\vartheta}}}%
  \def\bvarpi{{\bmath{\varpi}}}%
  \def\bvarrho{{\bmath{\varrho}}}%
  \def\bvarsigma{{\bmath{\varsigma}}}%
  \def\bvarphi{{\bmath{\varphi}}}%
  \def\baleph{{\bmath{\aleph}}}%
  \def\bimath{{\bmath{\imath}}}%
  \def\bjmath{{\bmath{\jmath}}}%
  \def\bell{{\bmath{\ell}}}%
  \def\bwp{{\bmath{\wp}}}%
  \def\bRe{{\bmath{\Re}}}%
  \def\bIm{{\bmath{\Im}}}%
  \def\bpartial{{\bmath{\partial}}}%
  \def\binfty{{\bmath{\infty}}}%
  \def\bprime{{\bmath{\prime}}}%
  \def\bemptyset{{\bmath{\emptyset}}}%
  \def\bnabla{{\bmath{\nabla}}}%
  \def\btop{{\bmath{\top}}}%
  \def\bbot{{\bmath{\bot}}}%
  \def\btriangle{{\bmath{\triangle}}}%
  \def\bforall{{\bmath{\forall}}}%
  \def\bexists{{\bmath{\exists}}}%
  \def\bneg{{\bmath{\neg}}}%
  \def\bflat{{\bmath{\flat}}}%
  \def\bnatural{{\bmath{\natural}}}%
  \def\bsharp{{\bmath{\sharp}}}%
  \def\bclubsuit{{\bmath{\clubsuit}}}%
  \def\bdiamondsuit{{\bmath{\diamondsuit}}}%
  \def\bheartsuit{{\bmath{\heartsuit}}}%
  \def\bspadesuit{{\bmath{\spadesuit}}}%
  \def\bsmallint{{\bmath{\smallint}}}%
  \def\btriangleleft{{\bmath{\triangleleft}}}%
  \def\btriangleright{{\bmath{\triangleright}}}%
  \def\bbigtriangleup{{\bmath{\bigtriangleup}}}%
  \def\bbigtriangledown{{\bmath{\bigtriangledown}}}%
  \def\bwedge{{\bmath{\wedge}}}%
  \def\bvee{{\bmath{\vee}}}%
  \def\bcap{{\bmath{\cap}}}%
  \def\bcup{{\bmath{\cup}}}%
  \def\bddagger{{\bmath{\ddagger}}}%
  \def\bdagger{{\bmath{\dagger}}}%
  \def\bsqcap{{\bmath{\sqcap}}}%
  \def\bsqcup{{\bmath{\sqcup}}}%
  \def\buplus{{\bmath{\uplus}}}%
  \def\bamalg{{\bmath{\amalg}}}%
  \def\bdiamond{{\bmath{\diamond}}}%
  \def\bbullet{{\bmath{\bullet}}}%
  \def\bwr{{\bmath{\wr}}}%
  \def\bdiv{{\bmath{\div}}}%
  \def\bodot{{\bmath{\odot}}}%
  \def\boslash{{\bmath{\oslash}}}%
  \def\botimes{{\bmath{\otimes}}}%
  \def\bominus{{\bmath{\ominus}}}%
  \def\boplus{{\bmath{\oplus}}}%
  \def\bmp{{\bmath{\mp}}}%
  \def\bpm{{\bmath{\pm}}}%
  \def\bcirc{{\bmath{\circ}}}%
  \def\bbigcirc{{\bmath{\bigcirc}}}%
  \def\bsetminus{{\bmath{\setminus}}}%
  \def\bcdot{{\bmath{\cdot}}}%
  \def\bast{{\bmath{\ast}}}%
  \def\btimes{{\bmath{\times}}}%
  \def\bstar{{\bmath{\star}}}%
  \def\bpropto{{\bmath{\propto}}}%
  \def\bsqsubseteq{{\bmath{\sqsubseteq}}}%
  \def\bsqsupseteq{{\bmath{\sqsupseteq}}}%
  \def\bparallel{{\bmath{\parallel}}}%
  \def\bmid{{\bmath{\mid}}}%
  \def\bdashv{{\bmath{\dashv}}}%
  \def\bvdash{{\bmath{\vdash}}}%
  \def\bnearrow{{\bmath{\nearrow}}}%
  \def\bsearrow{{\bmath{\searrow}}}%
  \def\bnwarrow{{\bmath{\nwarrow}}}%
  \def\bswarrow{{\bmath{\swarrow}}}%
  \def\bLeftrightarrow{{\bmath{\Leftrightarrow}}}%
  \def\bLeftarrow{{\bmath{\Leftarrow}}}%
  \def\bRightarrow{{\bmath{\Rightarrow}}}%
  \def\bleq{{\bmath{\leq}}}%
  \def\bgeq{{\bmath{\geq}}}%
  \def\bsucc{{\bmath{\succ}}}%
  \def\bprec{{\bmath{\prec}}}%
  \def\bapprox{{\bmath{\approx}}}%
  \def\bsucceq{{\bmath{\succeq}}}%
  \def\bpreceq{{\bmath{\preceq}}}%
  \def\bsupset{{\bmath{\supset}}}%
  \def\bsubset{{\bmath{\subset}}}%
  \def\bsupseteq{{\bmath{\supseteq}}}%
  \def\bsubseteq{{\bmath{\subseteq}}}%
  \def\bin{{\bmath{\in}}}%
  \def\bni{{\bmath{\ni}}}%
  \def\bgg{{\bmath{\gg}}}%
  \def\bll{{\bmath{\ll}}}%
  \def\bnot{{\bmath{\not}}}%
  \def\bleftrightarrow{{\bmath{\leftrightarrow}}}%
  \def\bleftarrow{{\bmath{\leftarrow}}}%
  \def\brightarrow{{\bmath{\rightarrow}}}%
  \def\bmapstochar{{\bmath{\mapstochar}}}%
  \def\bsim{{\bmath{\sim}}}%
  \def\bsimeq{{\bmath{\simeq}}}%
  \def\bperp{{\bmath{\perp}}}%
  \def\bequiv{{\bmath{\equiv}}}%
  \def\basymp{{\bmath{\asymp}}}%
  \def\bsmile{{\bmath{\smile}}}%
  \def\bfrown{{\bmath{\frown}}}%
  \def\bleftharpoonup{{\bmath{\leftharpoonup}}}%
  \def\bleftharpoondown{{\bmath{\leftharpoondown}}}%
  \def\brightharpoonup{{\bmath{\rightharpoonup}}}%
  \def\brightharpoondown{{\bmath{\rightharpoondown}}}%
  \def\blhook{{\bmath{\lhook}}}%
  \def\brhook{{\bmath{\rhook}}}%
  \def\bldotp{{\bmath{\ldotp}}}%
  \def\bcdotp{{\bmath{\cdotp}}}%
}

% Make \, work in non-math mode
\def\,{\relax\ifmmode \mskip\thinmuskip\else \thinspace\fi}
\let\protect=\relax

\long\def\@ifundefined#1#2#3{\expandafter\ifx\csname
  #1\endcsname\relax#2\else#3\fi}

%%%%%%%%%%%%%%%%%%%%%%%%%%%%%%%%%%%%%%%%%

% NewFont.sty: ALPHA VERSION patchlevel 8, 16th August 1994, M. Reed

% \addtom@thgroup{math font loading info}
% Adds to internal \math@groups definition, which is executed at the end
% of each size changing command. It is called by \NewSymbolFont.

\newtoks\math@groups \math@groups={}
\def\addtom@thgroup#1#2{#1\expandafter{\the#1#2}} %  \mac={new\the\mac}

% Make TeX change the values of \s@ze, \ss@ze, \sss@ze when \@npt is
% executed. This makes it possible for math characters to be loaded
% at the correct size automatically when the size is changed.

% \addtosizeh@ok{x}{10}{7}{5}

\def\addtosizeh@ok#1#2#3#4{%
  \expandafter\def\csname @#1pt\endcsname{%
    \def\s@ze{#2}\def\ss@ze{#3}\def\sss@ze{#4}\the\math@groups%
  }%
}

% \resetsizehook allows the size parameters to be reset after \addtosizeh@ok
% has been called (it re-defines \@npt).
% e.g JFM which requires \xpt to have 10.5pt instead of 10pt.
% Note: \resetsizehook must be used in the preamble BEFORE any
% \New... commands.

% e.g. \resetsizehook{x}{10.5}{7}{5}

\let\resetsizehook=\addtosizeh@ok

% Standard LaTeX sizes

\ifprod@font
%  \addtosizeh@ok{v}    {5} {5}  {5}
%  \addtosizeh@ok{vi}   {6} {6}  {6}
%  \addtosizeh@ok{vii}  {7} {6}  {5}
  \addtosizeh@ok{viii} {8} {6}  {5}
  \addtosizeh@ok{ix}   {9} {6}  {5}
  \addtosizeh@ok{x}    {10}{7}  {5}
  \addtosizeh@ok{xi}   {11}{8}  {6}
%  \addtosizeh@ok{xii}  {12}{8}  {6}
  \addtosizeh@ok{xiv}  {14}{10} {7}
  \addtosizeh@ok{xvii} {17}{12}{10}
%  \addtosizeh@ok{xx}   {20}{14}{12}
%  \addtosizeh@ok{xxv}  {25}{20}{17}
\else
%  \addtosizeh@ok{v}    {5}     {5}     {5}
%  \addtosizeh@ok{vi}   {6}     {6}     {6}
%  \addtosizeh@ok{vii}  {7}     {6}     {5}
  \addtosizeh@ok{viii} {8}     {6}     {5}
  \addtosizeh@ok{ix}   {9}     {6}     {5}
  \addtosizeh@ok{x}    {10}    {7}     {5}
  \addtosizeh@ok{xi}   {10.95} {8}     {6}
%  \addtosizeh@ok{xii}  {12}    {8}     {6}
  \addtosizeh@ok{xiv}  {14.4}  {10}    {7}
  \addtosizeh@ok{xvii} {17.28} {12}    {10}
%  \addtosizeh@ok{xx}   {20.74} {14.4}  {12}
%  \addtosizeh@ok{xxv}  {24.88} {20.74} {17.28}
\fi

\def\get@font#1#2#3{%
  \edef\fonts@ze{\romannumeral#3}%         10 -> x
  \edef\fontn@me{\fonts@ze#1}%             AMSa -> xAMSa
  \@ifundefined{\fontn@me}%
    {%%\typeout{defining \fontn@me}%
     \global\expandafter\font\csname \fontn@me\endcsname=#2 at #3pt}%
    {}%
}

\def\ass@tfont#1#2{%
  \xdef\fam@name{\csname #1\endcsname}%
  \xdef\font@name{\csname #2\endcsname}%
  \let\textfont@name\font@name
  \textfont\fam@name\textfont@name
}

\def\ass@sfont#1#2{%
  \xdef\fam@name{\csname #1\endcsname}%
  \xdef\font@name{\csname #2\endcsname}%
  \let\textfont@name\font@name
  \scriptfont\fam@name\textfont@name
}

\def\ass@ssfont#1#2{%
  \xdef\fam@name{\csname #1\endcsname}%
  \xdef\font@name{\csname #2\endcsname}%
  \let\textfont@name\font@name
  \scriptscriptfont\fam@name\textfont@name
}

%                fam name  base font  (allocates a \newfam)
% \NewSymbolFont {AMSa}    {mtxm10}

\def\NewSymbolFont#1#2{%
  \expandafter\ifx\csname sym#1fam\endcsname\relax % if not defined
    \expandafter\newfam\csname sym#1fam\endcsname
    \expandafter\edef\csname sym#1fam\endcsname{\the\allocationnumber}%
    \addtom@thgroup\math@groups{%
      \get@font{#1}{#2}{\s@ze}%
      \ass@tfont{sym#1fam}{\fontn@me}%
      \get@font{#1}{#2}{\ss@ze}%
      \ass@sfont{sym#1fam}{\fontn@me}%
      \get@font{#1}{#2}{\sss@ze}%
      \ass@ssfont{sym#1fam}{\fontn@me}%
    }%
  \else
    \errmessage{Family `#1' already defined}%
  \fi
}

%                symbol         type fam    pos (hex)
% \NewMathSymbol {\blacksquare} {0}  {AMSa} {04}

\def\NewMathSymbol#1#2#3#4{%
  \edef\f@mly{\expandafter\hexnumber{\csname sym#3fam\endcsname}}%
  \mathchardef#1="#2\f@mly#4\relax
}

%                  macro name  type  fam1   pos  fam2   pos
% \NewMathDelimiter{\ulcorner} {4}   {AMSa} {70} {AMSb} {70}

\newif\ifd@f

\def\NewMathDelimiter#1#2#3#4#5#6{%
  \d@ftrue
  \expandafter\ifx\csname sym#3fam\endcsname\relax
    \d@ffalse \errmessage{Family `#3' is not defined}%
  \fi
  \expandafter\ifx\csname sym#5fam\endcsname\relax
    \d@ffalse \errmessage{Family `#5' is not defined}%
  \fi
  \ifd@f
    \edef\f@mly{\expandafter\hexnumber{\csname sym#3fam\endcsname}}%
    \edef\f@mlytw@{\expandafter\hexnumber{\csname sym#5fam\endcsname}}%
    \xdef#1{\delimiter"#2\f@mly #4\f@mlytw@ #6\relax}%
  \fi
}

%                  macro name  base font  skewchar setting e.g '60 (octal)
% \NewMathAlphabet {mathbssi}  {mtmisb10} {}

\def\setboxz@h{\setbox\z@\hbox}
\def\wdz@{\wd\z@}
\def\boxz@{\box\z@}
\def\setbox@ne{\setbox\@ne}
\def\wd@ne{\wd\@ne}

\def\math@atom#1#2{%
   \binrel@{#1}\binrel@@{#2}}
\def\binrel@#1{\setboxz@h{\thinmuskip0mu
  \medmuskip\m@ne mu\thickmuskip\@ne mu$#1\m@th$}%
 \setbox@ne\hbox{\thinmuskip0mu\medmuskip\m@ne mu\thickmuskip
  \@ne mu${}#1{}\m@th$}%
 \setbox\tw@\hbox{\hskip\wd@ne\hskip-\wdz@}}
\def\binrel@@#1{\ifdim\wd2<\z@\mathbin{#1}\else\ifdim\wd\tw@>\z@
 \mathrel{#1}\else{#1}\fi\fi}

\def\m@thit{1}

\def\set@skchar#1{\global\expandafter\skewchar
  \csname\fontn@me\endcsname=#1\relax}

\def\NewMathAlphabet#1#2#3{%
  \def\tst{#3}%
  \ifx\tst\empty\else % if a \skewchar setting is present
    \expandafter\gdef\csname #1@sc\endcsname{}%  \def\cmd@sc{..}
  \fi
  \expandafter\def\csname #1\endcsname{%  \def\cmd{\protect\@cmd}
    \protect\csname @#1\endcsname}%
  \expandafter\def\csname @#1\endcsname##1{%  \def\@cmd{..}
    {%
    \begingroup
      \get@font{#1}{#2}{\s@ze}%
      \@ifundefined{#1@sc}{}{\set@skchar{#3}}%
      \ass@tfont{m@thit}{\fontn@me}%
      \get@font{#1}{#2}{\ss@ze}%
      \@ifundefined{#1@sc}{}{\set@skchar{#3}}%
      \ass@sfont{m@thit}{\fontn@me}%
      \get@font{#1}{#2}{\sss@ze}%
      \@ifundefined{#1@sc}{}{\set@skchar{#3}}%
      \ass@ssfont{m@thit}{\fontn@me}%
      \math@atom{##1}{%
      \mathchoice%
        {\hbox{$\m@th\displaystyle##1$}}%
        {\hbox{$\m@th\textstyle##1$}}%
        {\hbox{$\m@th\scriptstyle##1$}}%
        {\hbox{$\m@th\scriptscriptstyle##1$}}}%
    \endgroup
    }%
  }%
}

%                  macro name  base font  hyphenchar setting e.g -1 (off)
% \NewTextAlphabet {textbfit}  {mtbxti10} {}

% save a family if \NewTextAlphabet is not used.
\newif\iffirstta  \firsttatrue

\def\set@hchar#1{\global\expandafter\hyphenchar
  \csname\fontn@me\endcsname=#1\relax}

\def\NewTextAlphabet#1#2#3{%
  \iffirstta
    \global\firsttafalse
    \newfam\scratchfam
    \edef\scrt@fam{\the\allocationnumber}%
  \fi
  \def\tst{#3}%
  \ifx\tst\empty\else % if a \hyphenchar setting is required
    \expandafter\gdef\csname #1@hc\endcsname{}%  \def\cmd@sc{..}
  \fi
  \expandafter\def\csname #1\endcsname{%  \def\cmd{\protect\t@cmd}
    \protect\csname t@#1\endcsname}%
  \long\expandafter\def\csname t@#1\endcsname##1{%  \def\t@cmd{..}
    \ifmmode
      \typeout{Warning: do not use \expandafter\string\csname #1\endcsname
        \space in math mode}\fi%
    {%
      \get@font{#1}{#2}{\s@ze}\let\t@xtfnt=\fontn@me\relax
      \@ifundefined{#1@hc}{}{\set@hchar{#3}}%
      \ass@tfont{scrt@fam}{\fontn@me}%
      \get@font{#1}{#2}{\ss@ze}%
      \@ifundefined{#1@hc}{}{\set@hchar{#3}}%
      \ass@sfont{scrt@fam}{\fontn@me}%
      \get@font{#1}{#2}{\sss@ze}%
      \@ifundefined{#1@hc}{}{\set@hchar{#3}}%
      \ass@ssfont{scrt@fam}{\fontn@me}%
      \fam\scratchfam\csname\t@xtfnt\endcsname
    ##1%
    }%
  }%
  \expandafter\def\csname #1shape%  \def\cmdshape{\protect\@cmdshape}
    \endcsname{\protect\csname @#1shape\endcsname}%
  \expandafter\def\csname @#1shape\endcsname{%  \def\@cmdshape
    \ifmmode
      \typeout{Warning: do not use \expandafter\string\csname
        #1shape\endcsname \space in math mode}\fi
      \get@font{#1}{#2}{\s@ze}\let\t@xtfnt=\fontn@me\relax
      \@ifundefined{#1@hc}{}{\set@hchar{#3}}%
      \ass@tfont{scrt@fam}{\fontn@me}%
      \get@font{#1}{#2}{\ss@ze}%
      \@ifundefined{#1@hc}{}{\set@hchar{#3}}%
      \ass@sfont{scrt@fam}{\fontn@me}%
      \get@font{#1}{#2}{\sss@ze}%
      \@ifundefined{#1@hc}{}{\set@hchar{#3}}%
      \ass@ssfont{scrt@fam}{\fontn@me}%
      \fam\scratchfam\csname\t@xtfnt\endcsname
  }%
}

% \bmath{math text}

\ifprod@font
  \def\math@itfnt{mtmib10}
  \def\math@syfnt{mtbsy10}
\else
  \def\math@itfnt{cmmib10}
  \def\math@syfnt{cmbsy10}
\fi

\def\m@thsy{2}

\def\bmath{\protect\@bmath}
\def\@bmath#1{%
  {%
  \begingroup
    \get@font{mthit}{\math@itfnt}{\s@ze}\set@skchar{'177}%
    \ass@tfont{m@thit}{\fontn@me}%
    \get@font{mthit}{\math@itfnt}{\ss@ze}\set@skchar{'177}%
    \ass@sfont{m@thit}{\fontn@me}%
    \get@font{mthit}{\math@itfnt}{\sss@ze}\set@skchar{'177}%
    \ass@ssfont{m@thit}{\fontn@me}%
    \get@font{mthsy}{\math@syfnt}{\s@ze}\set@skchar{'60}%
    \ass@tfont{m@thsy}{\fontn@me}%
    \get@font{mthsy}{\math@syfnt}{\ss@ze}\set@skchar{'60}%
    \ass@sfont{m@thsy}{\fontn@me}%
    \get@font{mthsy}{\math@syfnt}{\sss@ze}\set@skchar{'60}%
    \ass@ssfont{m@thsy}{\fontn@me}%
    \math@atom{#1}{%
    \mathchoice%
      {\hbox{$\m@th\displaystyle#1$}}%
      {\hbox{$\m@th\textstyle#1$}}%
      {\hbox{$\m@th\scriptstyle#1$}}%
      {\hbox{$\m@th\scriptscriptstyle#1$}}}%
  \endgroup
  }%
}

%%%%%%%%%%%%%%%%%%%%%%%%%%%%%%%%%%%%%%%%%

% Astronomy and Astrophysics symbol macros

\def\diameter{{\ifmmode\mathchoice
{\ooalign{\hfil\hbox{$\displaystyle/$}\hfil\crcr
{\hbox{$\displaystyle\mathchar"20D$}}}}
{\ooalign{\hfil\hbox{$\textstyle/$}\hfil\crcr
{\hbox{$\textstyle\mathchar"20D$}}}}
{\ooalign{\hfil\hbox{$\scriptstyle/$}\hfil\crcr
{\hbox{$\scriptstyle\mathchar"20D$}}}}
{\ooalign{\hfil\hbox{$\scriptscriptstyle/$}\hfil\crcr
{\hbox{$\scriptscriptstyle\mathchar"20D$}}}}
\else{\ooalign{\hfil/\hfil\crcr\mathhexbox20D}}%
\fi}}

\def\sq{\ifmmode\squareforqed\else{\unskip\nobreak\hfil
\penalty50\hskip1em\null\nobreak\hfil\squareforqed
\parfillskip=0pt\finalhyphendemerits=0\endgraf}\fi}
\def\squareforqed{\hbox{\rlap{$\sqcap$}$\sqcup$}}

\def\arcsec{\hbox{$^{\prime\prime}$}}

% Simulated Blackboard Bold symbols

\def\bbbc{{\mathchoice {\setbox0=\hbox{$\displaystyle\rm C$}\hbox{\hbox
to0pt{\kern0.4\wd0\vrule height0.9\ht0\hss}\box0}}
{\setbox0=\hbox{$\textstyle\rm C$}\hbox{\hbox
to0pt{\kern0.4\wd0\vrule height0.9\ht0\hss}\box0}}
{\setbox0=\hbox{$\scriptstyle\rm C$}\hbox{\hbox
to0pt{\kern0.4\wd0\vrule height0.9\ht0\hss}\box0}}
{\setbox0=\hbox{$\scriptscriptstyle\rm C$}\hbox{\hbox
to0pt{\kern0.4\wd0\vrule height0.9\ht0\hss}\box0}}}}
\def\bbbq{{\mathchoice {\setbox0=\hbox{$\displaystyle\rm
Q$}\hbox{\raise
0.15\ht0\hbox to0pt{\kern0.4\wd0\vrule height0.8\ht0\hss}\box0}}
{\setbox0=\hbox{$\textstyle\rm Q$}\hbox{\raise
0.15\ht0\hbox to0pt{\kern0.4\wd0\vrule height0.8\ht0\hss}\box0}}
{\setbox0=\hbox{$\scriptstyle\rm Q$}\hbox{\raise
0.15\ht0\hbox to0pt{\kern0.4\wd0\vrule height0.7\ht0\hss}\box0}}
{\setbox0=\hbox{$\scriptscriptstyle\rm Q$}\hbox{\raise
0.15\ht0\hbox to0pt{\kern0.4\wd0\vrule height0.7\ht0\hss}\box0}}}}
\def\bbbt{{\mathchoice {\setbox0=\hbox{$\displaystyle\rm
T$}\hbox{\hbox to0pt{\kern0.3\wd0\vrule height0.9\ht0\hss}\box0}}
{\setbox0=\hbox{$\textstyle\rm T$}\hbox{\hbox
to0pt{\kern0.3\wd0\vrule height0.9\ht0\hss}\box0}}
{\setbox0=\hbox{$\scriptstyle\rm T$}\hbox{\hbox
to0pt{\kern0.3\wd0\vrule height0.9\ht0\hss}\box0}}
{\setbox0=\hbox{$\scriptscriptstyle\rm T$}\hbox{\hbox
to0pt{\kern0.3\wd0\vrule height0.9\ht0\hss}\box0}}}}
\def\bbbs{{\mathchoice
{\setbox0=\hbox{$\displaystyle     \rm S$}\hbox{\raise0.5\ht0\hbox
to0pt{\kern0.35\wd0\vrule height0.45\ht0\hss}\hbox
to0pt{\kern0.55\wd0\vrule height0.5\ht0\hss}\box0}}
{\setbox0=\hbox{$\textstyle        \rm S$}\hbox{\raise0.5\ht0\hbox
to0pt{\kern0.35\wd0\vrule height0.45\ht0\hss}\hbox
to0pt{\kern0.55\wd0\vrule height0.5\ht0\hss}\box0}}
{\setbox0=\hbox{$\scriptstyle      \rm S$}\hbox{\raise0.5\ht0\hbox
to0pt{\kern0.35\wd0\vrule height0.45\ht0\hss}\raise0.05\ht0\hbox
to0pt{\kern0.5\wd0\vrule height0.45\ht0\hss}\box0}}
{\setbox0=\hbox{$\scriptscriptstyle\rm S$}\hbox{\raise0.5\ht0\hbox
to0pt{\kern0.4\wd0\vrule height0.45\ht0\hss}\raise0.05\ht0\hbox
to0pt{\kern0.55\wd0\vrule height0.45\ht0\hss}\box0}}}}
\def\bbbz{{\mathchoice {\hbox{$\sf\textstyle Z\kern-0.4em Z$}}
{\hbox{$\sf\textstyle Z\kern-0.4em Z$}}
{\hbox{$\sf\scriptstyle Z\kern-0.3em Z$}}
{\hbox{$\sf\scriptscriptstyle Z\kern-0.2em Z$}}}}

% NUMBER THE DESIGN ELEMENTS

\def\Nulle{0} % null element
\def\Afe{1}   % author affiliation
\def\Hae{2}   % heading A
\def\Hbe{3}   % heading B
\def\Hce{4}   % heading C
\def\Hde{5}   % heading D

% TEMPORARY REGISTERS

\newcount\LastMac       \LastMac=\Nulle

\newskip\half      \half=5.5pt plus 1.5pt minus 2.25pt
\newskip\one       \one=11pt plus 3pt minus 5.5pt
\newskip\onehalf   \onehalf=16.5pt plus 5.5pt minus 8.25pt
\newskip\two       \two=22pt plus 5.5pt minus 11pt

\def\Half{\addvspace{\half}}
\def\One{\addvspace{\one}}
\def\OneHalf{\addvspace{\onehalf}}
\def\Two{\addvspace{\two}}

\def\Raggedright{% set lines unjustified
  \rightskip=\z@ plus \hsize\relax
}

\def\Fullout{% set lines justified
  \rightskip=\z@\relax
}

\def\Hang#1#2{% set hanging indentation
  \hangindent=#1%
  \hangafter=#2\relax
}

% Pagestyles

\newif\ifsp@page
\def\pagestyle#1{\csname ps@#1\endcsname}
\def\thispagestyle#1{\global\sp@pagetrue\gdef\sp@type{#1}}

\def\ps@titlepage{%
  \def\@oddhead{\eightpoint\noindent \the\CatchLine
    \ifprod@font\else\qquad Printed\ \today\qquad
      (MN plain \TeX\ macros\ v\@version)\fi \hfil}%
  \let\@evenhead=\@oddhead
}

\def\ps@headings{%
  \def\@oddhead{\elevenpoint\it\noindent
    \hfill\the\RightHeader\hskip1.5em\rm\folio}%
  \def\@evenhead{\elevenpoint\noindent
    \folio\hskip1.5em\it\the\LeftHeader\hfill}%
}

\def\ps@plate{%
  \def\@oddhead{\eightpoint\noindent\plt@cap\hfil}%
  \def\@evenhead{\eightpoint\noindent\plt@cap\hfil}%
}

% DESIGN ELEMENT DEFINITIONS

% Article opening

\def\title#1{% article title
  \bgroup
    \vbox to 8pt{\vss}%
    \seventeenpoint
    \Raggedright
    \noindent \strut{\bf #1}\par
  \egroup
}

\def\author#1{% article author(s)
  \bgroup
    \ifnum\LastMac=\Afe \OneHalf\else \vskip 21pt\fi
    \fourteenpoint
    \Raggedright
    \noindent \strut #1\par
    \vskip 3pt%
  \egroup
}

\def\affiliation#1{% author(s) affiliation
  \bgroup
    \vskip -4pt%
    \eightpoint
    \Raggedright
    \noindent \strut {\it #1}\par
  \egroup
  \LastMac=\Afe\relax
}

\def\acceptedline#1{% acceptance date
  \bgroup
    \Two
    \eightpoint
    \Raggedright
    \noindent \strut #1\par
  \egroup
}

\long\def\abstract#1{%
  \bgroup
    \vskip 20pt%
    \everypar{\Hang{11pc}{0}}%
    \noindent{\ninebf ABSTRACT}\par
    \tenpoint
    \Fullout
    \noindent #1\par
  \egroup
}

\long\def\keywords#1{% keywords
  \bgroup
    \Half
    \everypar{\Hang{11pc}{0}}%
    \tenpoint
    \Fullout
    \noindent\hbox{\bf Key words:}\ #1\par
  \egroup
}

% The \maketitle macro ensures that the two spanning material appears
% at the top of the first page, and that it has two lines of space
% underneath it. If you forget this in you input, no output will be produced.
% The \BeginOpening (alias \begintopmatter) macro should be called at the
% very start of the input file, so that it is in force when the document
% starts. This ensures that when \maketitle is called that the group is
% closed, and the material gets printed.

\def\maketitle{%
  \EndOpening
  \ifsinglecol \else \MakePage\fi
}

% Page offset

\def\pageoffset#1#2{\hoffset=#1\relax\voffset=#2\relax}

% Counter setup

\def\@nameuse#1{\csname #1\endcsname}
\def\arabic#1{\@arabic{\@nameuse{#1}}}
\def\alph#1{\@alph{\@nameuse{#1}}}
\def\Alph#1{\@Alph{\@nameuse{#1}}}
\def\@arabic#1{\number #1}
\def\@Alph#1{\ifcase#1\or A\or B\or C\or D\else\@Ialph{#1}\fi}
\def\@Ialph#1{\ifcase#1\or \or \or \or \or E\or F\or G\or H\or I\or J\or
   K\or L\or M\or N\or O\or P\or Q\or R\or S\or T\or U\or V\or W\or X\or
   Y\or Z\else\errmessage{Counter out of range}\fi}
\def\@alph#1{\ifcase#1\or a\or b\or c\or d\else\@ialph{#1}\fi}
\def\@ialph#1{\ifcase#1\or \or \or \or \or e\or f\or g\or h\or i\or j\or
   k\or l\or m\or n\or o\or p\or q\or r\or s\or t\or u\or v\or w\or x\or y\or
   z\else\errmessage{Counter out of range}\fi}

% Equation auto-numbering

\newcount\Eqnno
\newcount\SubEqnno

\def\theeq{\arabic{Eqnno}}
\def\thesubeq{\alph{SubEqnno}}

\def\stepeq{\relax
  \global\SubEqnno \z@
  \global\advance\Eqnno \@ne\relax
  {\rm (\theeq)}%
}

\def\startsubeq{\relax
  \global\SubEqnno \z@
  \global\advance\Eqnno \@ne\relax
  \stepsubeq
}

\def\stepsubeq{\relax
  \global\advance\SubEqnno \@ne\relax
  {\rm (\theeq\thesubeq)}%
}

% Headings

\newcount\Sec        %  heading auto number counters
\newcount\SecSec
\newcount\SecSecSec

\def\thesection{\arabic{Sec}}
\def\thesubsection{\thesection.\arabic{SecSec}}
\def\thesubsubsection{\thesubsection.\arabic{SecSecSec}}

\Sec=\z@

\def\:{\let\@sptoken= } \:  % this makes \@sptoken a space token 
\def\:{\@xifnch} \expandafter\def\: {\futurelet\@tempc\@ifnch}

\def\@ifnextchar#1#2#3{%
  \let\@tempMACe #1%
  \def\@tempMACa{#2}%
  \def\@tempMACb{#3}%
  \futurelet \@tempMACc\@ifnch%
}

\def\@ifnch{%
\ifx \@tempMACc \@sptoken%
  \let\@tempMACd\@xifnch%
\else%
  \ifx \@tempMACc \@tempMACe%
    \let\@tempMACd\@tempMACa%
  \else%
    \let\@tempMACd\@tempMACb%
  \fi%
\fi%
\@tempMACd%
}

\def\@ifstar#1#2{\@ifnextchar *{\def\@tempMACa*{#1}\@tempMACa}{#2}}

\newskip\@tempskipb

\def\addvspace#1{%
  \ifvmode\else \endgraf\fi%
  \ifdim\lastskip=\z@%
    \vskip #1\relax%
  \else%
    \@tempskipb#1\relax\@xaddvskip%
  \fi%
}

\def\@xaddvskip{%
  \ifdim\lastskip<\@tempskipb%
    \vskip-\lastskip%
    \vskip\@tempskipb\relax%
  \else%
    \ifdim\@tempskipb<\z@%
      \ifdim\lastskip<\z@ \else%
        \advance\@tempskipb\lastskip%
        \vskip-\lastskip\vskip\@tempskipb%
      \fi%
    \fi%
  \fi%
}

\newskip\@tmpSKIP

\def\addpen#1{%
  \ifvmode
    \if@nobreak
    \else
      \ifdim\lastskip=\z@
        \penalty#1\relax
      \else
        \@tmpSKIP=\lastskip
        \vskip -\lastskip
        \penalty#1\vskip\@tmpSKIP
      \fi
    \fi
  \fi
}

\newcount\@clubpen   \@clubpen=\clubpenalty
\newif\if@nobreak    \@nobreakfalse

\def\@noafterindent{%
  \global\@nobreaktrue
  \everypar{\if@nobreak
              \global\@nobreakfalse
              \clubpenalty \@M
              {\setbox\z@\lastbox}%
              \LastMac=\Nulle\relax%
            \else
              \clubpenalty \@clubpen
              \everypar{}%
            \fi}%
}

\newcount\gds@cbrk   \gds@cbrk=-300

\def\@nohdbrk{\interlinepenalty \@M\relax}

\let\@par=\par
\def\@restorepar{\def\par{\@par}}

\newif\if@endpe   \@endpefalse
 
\def\@doendpe{\@endpetrue \@nobreakfalse \LastMac=\Nulle\relax%
     \def\par{\@restorepar\everypar{}\par\@endpefalse}%
              \everypar{\setbox\z@\lastbox\everypar{}\@endpefalse}%
}

\def\section{\@ifstar{\@ssection}{\@section}}

\def\@section#1{% heading A (\section{....})
  \if@nobreak
    \everypar{}%
    \ifnum\LastMac=\Hae \addvspace{\half}\fi
  \else
    \addpen{\gds@cbrk}%
    \addvspace{\two}%
  \fi
  \bgroup
    \ninepoint\bf
    \Raggedright
    \global\advance\Sec \@ne
    \ifappendix
      \global\Eqnno=\z@ \global\SubEqnno=\z@\relax
      \def\ch@ck{#1}%
      \ifx\ch@ck\empty \def\c@lon{}\else\def\c@lon{:}\fi
      \noindent\@nohdbrk APPENDIX\ \thesection\c@lon\hskip 0.5em%
        \uppercase{#1}\par
    \else
      \noindent\@nohdbrk\thesection\hskip 1pc \uppercase{#1}\par
    \fi
    \global\SecSec=\z@
  \egroup
  \nobreak
  \vskip\half
  \nobreak
  \@noafterindent
  \LastMac=\Hae\relax
}

\def\@ssection#1{%  main section heading (\section*{....})
  \if@nobreak
    \everypar{}%
    \ifnum\LastMac=\Hae \addvspace{\half}\fi
  \else
    \addpen{\gds@cbrk}%
    \addvspace{\two}%
  \fi
  \bgroup
    \ninepoint\bf
    \Raggedright
%    \ifappendix
%      \global\Eqnno=\z@ \global\SubEqnno=\z@\relax % mh in apps dont reset
%      \noindent\@nohdbrk APPENDIX:\hskip 0.5em%
%        \uppercase{#1}\par
%    \else
    \noindent\@nohdbrk\uppercase{#1}\par
%    \fi
  \egroup
  \nobreak
  \vskip\half
  \nobreak
  \@noafterindent
  \LastMac=\Hae\relax
}

\def\subsection{\@ifstar{\@ssubsection}{\@subsection}}

\def\@subsection#1{% heading B
  \if@nobreak
    \everypar{}%
    \ifnum\LastMac=\Hae \addvspace{1pt plus 1pt minus .5pt}\fi
  \else
    \addpen{\gds@cbrk}%
    \addvspace{\onehalf}%
  \fi
  \bgroup
    \ninepoint\bf
    \Raggedright
    \global\advance\SecSec \@ne
    \noindent\@nohdbrk\thesubsection \hskip 1pc\relax #1\par
    \global\SecSecSec=\z@
  \egroup
  \nobreak
  \vskip\half
  \nobreak
  \@noafterindent
  \LastMac=\Hbe\relax
}

\def\@ssubsection#1{% heading B*
  \if@nobreak
    \everypar{}%
    \ifnum\LastMac=\Hae \addvspace{1pt plus 1pt minus .5pt}\fi
  \else
    \addpen{\gds@cbrk}%
    \addvspace{\onehalf}%
  \fi
  \bgroup
    \ninepoint\bf
    \Raggedright
    \noindent\@nohdbrk #1\par
  \egroup
  \nobreak
  \vskip\half
  \nobreak
  \@noafterindent
  \LastMac=\Hbe\relax
}

\def\subsubsection{\@ifstar{\@ssubsubsection}{\@subsubsection}}

\def\@subsubsection#1{% heading C
  \if@nobreak
    \everypar{}%
    \ifnum\LastMac=\Hbe \addvspace{1pt plus 1pt minus .5pt}\fi
  \else
    \addpen{\gds@cbrk}%
    \addvspace{\onehalf}%
  \fi
  \bgroup
    \ninepoint\it
    \Raggedright
    \global\advance\SecSecSec \@ne
    \noindent\@nohdbrk\thesubsubsection \hskip 1pc\relax #1\par
  \egroup
  \nobreak
  \vskip\half
  \nobreak
  \@noafterindent
  \LastMac=\Hce\relax
}

\def\@ssubsubsection#1{% heading C*
  \if@nobreak
    \everypar{}%
    \ifnum\LastMac=\Hbe \addvspace{1pt plus 1pt minus .5pt}\fi
  \else
    \addpen{\gds@cbrk}%
    \addvspace{\onehalf}%
  \fi
  \bgroup
    \ninepoint\it
    \Raggedright
    \noindent\@nohdbrk #1\par
  \egroup
  \nobreak
  \vskip\half
  \nobreak
  \@noafterindent
  \LastMac=\Hce\relax
}

\def\paragraph#1{% heading D
  \if@nobreak
    \everypar{}%
  \else
    \addpen{\gds@cbrk}%
    \addvspace{\one}%
  \fi%
  \bgroup%
    \ninepoint\it
    \noindent #1\ \nobreak%
  \egroup
  \LastMac=\Hde\relax
  \ignorespaces
}

% Appendix

\newif\ifappendix

\def\appendix{%
  \global\appendixtrue
  \def\thesection{\Alph{Sec}}%
  \def\thesubsection{\thesection\arabic{SecSec}}%
  \def\theeq{\thesection\arabic{Eqnno}}%
  \Sec=\z@ \SecSec=\z@ \SecSecSec=\z@ \Eqnno=\z@ \SubEqnno=\z@\relax
}

% Text

 % provided for backward compatibility

% Lists

\def\beginlist{%
  \par\if@nobreak \else\addvspace{\half}\fi%
  \bgroup%
    \ninepoint
    \let\item=\list@item%
}

\def\list@item{%
  \par\noindent\hskip 1em\relax%
  \ignorespaces%
}

\def\endlist{\par\egroup\addvspace{\half}\@doendpe}

% References

\def\beginrefs{%
  \par
  \bgroup
    \eightpoint
    \Fullout
    \let\bibitem=\bib@item
}

\def\bib@item{%
  \par\parindent=1.5em\Hang{1.5em}{1}%
  \everypar={\Hang{1.5em}{1}\ignorespaces}%
  \noindent\ignorespaces
}

\def\endrefs{\par\egroup\@doendpe}

% Page heads

\newtoks\CatchLine

\def\@journal{Mon.\ Not.\ R.\ Astron.\ Soc.\ }  % The journal title string
\def\@pubyear{1994}        % Assign a default publication year
\def\@pagerange{000--000}  % Assign a default page-range
\def\@volume{000}          % Assign a default volume number
\def\@microfiche{}         %

\def\pubyear#1{\gdef\@pubyear{#1}\@makecatchline}
\def\pagerange#1{\gdef\@pagerange{#1}\@makecatchline}
\def\volume#1{\gdef\@volume{#1}\@makecatchline}
\def\microfiche#1{\gdef\@microfiche{and Microfiche\ #1}\@makecatchline}

\def\@makecatchline{%
  \global\CatchLine{%
    {\rm \@journal {\bf \@volume},\ \@pagerange\ (\@pubyear)\ \@microfiche}}%
}

\@makecatchline % Assign a catchline, using the above defaults

\newtoks\LeftHeader
\def\shortauthor#1{% left page head
  \global\LeftHeader{#1}%
}

\newtoks\RightHeader
\def\shorttitle#1{% right page head
  \global\RightHeader{#1}%
}

\def\PageHead{% recto/verso running heads
  \begingroup
    \ifsp@page
      \csname ps@\sp@type\endcsname
      \global\sp@pagefalse
    \fi
    \ifodd\pageno
      \let\the@head=\@oddhead
    \else
      \let\the@head=\@evenhead
    \fi
    \vbox to \z@{\vskip-22.5\p@%
      \hbox to \PageWidth{\vbox to8.5\p@{}%
        \the@head
      }%
    \vss}%
  \endgroup
  \nointerlineskip
}

\def\today{%
  \number\day\space
  \ifcase\month\or January\or February\or March\or April\or May\or June\or
    July\or August\or September\or October\or November\or December\fi
  \space\number\year%
}

\def\PageFoot{} % No page footer as default

\def\authorcomment#1{%
  \gdef\PageFoot{%
    \nointerlineskip%
    \vbox to 22pt{\vfil%
      \hbox to \PageWidth{\elevenpoint\noindent \hfil #1 \hfil}}%
  }%
}

% Plate pages

\newif\ifplate@page
\newbox\plt@box

\def\beginplatepage{%
  \let\plate=\plate@head
  \let\caption=\fig@caption
  \global\setbox\plt@box=\vbox\bgroup
  \TEMPDIMEN=\PageWidth % For \fig@caption test
  \hsize=\PageWidth\relax
}

\def\endplatepage{\par\egroup\global\plate@pagetrue}
\def\plate@head#1{\gdef\plt@cap{#1}}

% Letters option

\def\letters{%
  \gdef\folio{\ifnum\pageno<\z@ L\romannumeral-\pageno
    \else L\number\pageno \fi}%
}

% Math setup

% The standard math indentation
\newdimen\mathindent

\global\mathindent=\z@
\global\everydisplay{\global\@dspwd=\displaywidth\displaysetup}

% New versions of \displaylines, \eqalign, \eqalignno for
% when non-centered math is in use.

\def\@displaylines#1{% (for non-centered math)
  {}$\displ@y\hbox{\vbox{\halign{$\@lign\hfil\displaystyle##\hfil$\crcr
  #1\crcr}}}${}%
}

\def\@eqalign#1{\null\vcenter{\openup\jot\m@th% (for non-centered math)
  \ialign{\strut\hfil$\displaystyle{##}$&$\displaystyle{{}##}$\hfil
      \crcr#1\crcr}}%
}

\def\@eqalignno#1{% (for non-centered math)
  \global\advance\@dspwd by -\mathindent%
  {}$\displ@y\hbox{\vbox{\halign to\@dspwd%
  {\hfil$\@lign\displaystyle{##}$\tabskip\z@skip
  &$\@lign\displaystyle{{}##}$\hfil\tabskip\centering
  &\llap{$\@lign##$}\tabskip\z@skip\crcr
  #1\crcr}}}${}%
}

% When equations are flushleft ensure, that \displaylines,
% \eqalign, \eqalignno and \leqalignno point to the new versions of
% the macros. Also make \leqalignno act like \eqalignno, otherwise the
% equation text would `crash' into the equation number.

\global\let\displaylines=\@displaylines
\global\let\eqalign=\@eqalign
\global\let\eqalignno=\@eqalignno
\global\let\leqalignno=\@eqalignno

\newdimen\@dspwd   \@dspwd=\z@
\newif\if@eqno
\newif\if@leqno
\newtoks\@eqn
\newtoks\@eq

\def\displaysetup#1$${\displaytest#1\eqno\eqno\displaytest}

\def\displaytest#1\eqno#2\eqno#3\displaytest{%
 \if!#3!\ldisplaytest#1\leqno\leqno\ldisplaytest
 \else\@eqnotrue\@leqnofalse\@eqn={#2}\@eq={#1}\fi
 \generaldisplay$$}

\def\ldisplaytest#1\leqno#2\leqno#3\ldisplaytest{%
\@eq={#1}%
 \if!#3!\@eqnofalse\else\@eqnotrue\@leqnotrue
  \@eqn={#2}\fi}

\def\generaldisplay{%
  \if@eqno
    \if@leqno
      \hbox to \displaywidth{\noindent
        \rlap{$\displaystyle\the\@eqn$}%
        \hskip\mathindent$\displaystyle\the\@eq$\hfil}%
    \else
      \hbox to \displaywidth{\noindent
        \hskip\mathindent
        $\displaystyle\the\@eq$\hfil$\displaystyle\the\@eqn$}%
    \fi
  \else
    \hbox to \displaywidth{\noindent
      \hskip\mathindent$\displaystyle\the\@eq$\hfil}%
  \fi
}

% Finishing notice

\def\@notice{%
  \par\Two%
  \noindent{\b@ls{11pt}\ninerm This paper has been produced using the
    Royal Astronomical Society/Blackwell Science \TeX\ macros.\par}%
}

% redefine \bye to output our identification notice :
\outer\def\bye{\@notice\par\vfill\supereject\end}

% define a sign on :

\def\start@mess{%
  Monthly notices of the RAS journal style (\@typeface)\space
    v\@version,\space \@verdate.%
}

\everyjob{\Warn{\start@mess}}

% Two-column macros

%--------------------------------------------------------%
%                     INITIALISATION                     %
%--------------------------------------------------------%

\newif\if@debug \@debugfalse  %  when false, only warnings displayed

\def\Print#1{\if@debug\immediate\write16{#1}\else \fi}
\def\Warn#1{\immediate\write16{#1}}
\def\wlog#1{}

\newcount\Iteration % temporary loop counter

\def\Single{0} \def\Double{1}                 % ItemSPAN's
\def\Figure{0} \def\Table{1}                  % ItemTYPE's

\def\InStack{0}  % ItemSTATUS
\def\InZoneA{1}
\def\InZoneB{2}
\def\InZoneC{3}

\newcount\TEMPCOUNT % temporary count register
\newdimen\TEMPDIMEN % temporary dimen register
\newbox\TEMPBOX     % temporary box register
\newbox\VOIDBOX     % a box which is permenately void

\newcount\LengthOfStack % number of items currently in stack
\newcount\MaxItems      % maximum number of items allowed in stack
\newcount\StackPointer
\newcount\Point         % used in calculation for generating the
                        % physical address of an item in the stack.
\newcount\NextFigure    % number of next figure to be output
\newcount\NextTable     % number of next table to be output
\newcount\NextItem      % Next item to consider by order in stack

\newcount\StatusStack   % set to point to top of STATUS stack
\newcount\NumStack      % set to point to top of NUMBER stack
\newcount\TypeStack     % set to point to top of TYPE stack
\newcount\SpanStack     % set to point to top of SPAN stack
\newcount\BoxStack      % set to point to top of BOX stack

\newcount\ItemSTATUS    % status of present item
\newcount\ItemNUMBER    % number of present item
\newcount\ItemTYPE      % type of present item
\newcount\ItemSPAN      % span of present item
\newbox\ItemBOX         % box of present item
\newdimen\ItemSIZE      % size of present item
                        % (calculated by GetItemBOX)

\newdimen\PageHeight    % vertical measure of full page
\newdimen\TextLeading   % distance between baselines of body text
\newdimen\Feathering    % amount of interline stretch
\newcount\LinesPerPage  % height of page in text lines
\newdimen\ColumnWidth   % width of 1 column of text
\newdimen\ColumnGap     % size of gap between columns
\newdimen\PageWidth     % = \ColumnWidth * 2 + \ColumnGap
\newdimen\BodgeHeight   % Bodge to align figures and tables with text
\newcount\Leading       % Set to same as \TextLeading above

\newdimen\ZoneBSize  % size of items in ZoneB
\newdimen\TextSize   % size of text in ZoneB
\newbox\ZoneABOX     % contains Zone A material
\newbox\ZoneBBOX     % contains Zone B material
\newbox\ZoneCBOX     % contains Zone C material

\newif\ifFirstSingleItem
\newif\ifFirstZoneA
\newif\ifMakePageInComplete
\newif\ifMoreFigures \MoreFiguresfalse % set true in join stack
\newif\ifMoreTables  \MoreTablesfalse  % set true in join stack

\newif\ifFigInZoneB % used to determine in which zone an item
\newif\ifFigInZoneC % will be placed based on what is in other
\newif\ifTabInZoneB % zones already for a given page.
\newif\ifTabInZoneC

\newif\ifZoneAFullPage

\newbox\MidBOX    % = LeftBOX+gap+RightBOX
\newbox\LeftBOX
\newbox\RightBOX
\newbox\PageBOX   % complete made-up page

\newif\ifLeftCOL  % flags first pass through output routine
\LeftCOLtrue

\newdimen\ZoneBAdjust

\newcount\ItemFits
\def\Yes{1}
\def\No{2}

% Setup file.

\MaxItems=15
\NextFigure=\z@        % used for article opening
\NextTable=\@ne

\BodgeHeight=6pt
\TextLeading=11pt    % baselineskip of body text
\Leading=11
\Feathering=\z@      % amount of interline stretch
\LinesPerPage=61     % number of text lines per full page -1
\topskip=\TextLeading
\ColumnWidth=20pc    % width of text columns
\ColumnGap=2pc       % gap between columns

\newskip\ItemSepamount  % space between floats
\ItemSepamount=\TextLeading plus \TextLeading minus 4pt

\parskip=\z@ plus .1pt
\parindent=18pt
\widowpenalty=\z@
\clubpenalty=10000
\tolerance=1500
\hbadness=1500
\abovedisplayskip=6pt plus 2pt minus 1pt
\belowdisplayskip=6pt plus 2pt minus 1pt
\abovedisplayshortskip=6pt plus 2pt minus 1pt
\belowdisplayshortskip=6pt plus 2pt minus 1pt

\frenchspacing

\ninepoint % start main text size

\PageHeight=682pt
\PageWidth=2\ColumnWidth
\advance\PageWidth by \ColumnGap

\pagestyle{headings}

%--------------------------------------------------------%
%                         STACKS                         %
%--------------------------------------------------------%

% THE ITEM STACK
% The item stack contains contains figures and tables
% in the order in which they appear in the article source
% code.

% allocate stack space

\newcount\DUMMY \StatusStack=\allocationnumber
\newcount\DUMMY \newcount\DUMMY \newcount\DUMMY 
\newcount\DUMMY \newcount\DUMMY \newcount\DUMMY 
\newcount\DUMMY \newcount\DUMMY \newcount\DUMMY
\newcount\DUMMY \newcount\DUMMY \newcount\DUMMY 
\newcount\DUMMY \newcount\DUMMY \newcount\DUMMY

\newcount\DUMMY \NumStack=\allocationnumber
\newcount\DUMMY \newcount\DUMMY \newcount\DUMMY 
\newcount\DUMMY \newcount\DUMMY \newcount\DUMMY 
\newcount\DUMMY \newcount\DUMMY \newcount\DUMMY 
\newcount\DUMMY \newcount\DUMMY \newcount\DUMMY 
\newcount\DUMMY \newcount\DUMMY \newcount\DUMMY

\newcount\DUMMY \TypeStack=\allocationnumber
\newcount\DUMMY \newcount\DUMMY \newcount\DUMMY 
\newcount\DUMMY \newcount\DUMMY \newcount\DUMMY 
\newcount\DUMMY \newcount\DUMMY \newcount\DUMMY 
\newcount\DUMMY \newcount\DUMMY \newcount\DUMMY 
\newcount\DUMMY \newcount\DUMMY \newcount\DUMMY

\newcount\DUMMY \SpanStack=\allocationnumber
\newcount\DUMMY \newcount\DUMMY \newcount\DUMMY 
\newcount\DUMMY \newcount\DUMMY \newcount\DUMMY 
\newcount\DUMMY \newcount\DUMMY \newcount\DUMMY 
\newcount\DUMMY \newcount\DUMMY \newcount\DUMMY 
\newcount\DUMMY \newcount\DUMMY \newcount\DUMMY

\newbox\DUMMY   \BoxStack=\allocationnumber
\newbox\DUMMY   \newbox\DUMMY \newbox\DUMMY 
\newbox\DUMMY   \newbox\DUMMY \newbox\DUMMY 
\newbox\DUMMY   \newbox\DUMMY \newbox\DUMMY 
\newbox\DUMMY   \newbox\DUMMY \newbox\DUMMY 
\newbox\DUMMY   \newbox\DUMMY \newbox\DUMMY

\def\wlog{\immediate\write\m@ne}

% \GetItemSTATUS, \GetItemNUMBER, \GetItemTYPE, \GetItemSPAN,
% \GetItemBox 
% are used to get details of a particular item from the item
% stack. The argument to each of these is the items position
% in the stack (usually \StackPointer)...not the items number.

\def\GetItemAll#1{%
 \GetItemSTATUS{#1}
 \GetItemNUMBER{#1}
 \GetItemTYPE{#1}
 \GetItemSPAN{#1}
 \GetItemBOX{#1}
}

% Note: \LeaveStack uses this routine. Do not destroy \Point
\def\GetItemSTATUS#1{%
 \Point=\StatusStack
 \advance\Point by #1
 \global\ItemSTATUS=\count\Point
}

% Note: \LeaveStack uses this routine. Do not destroy \Point
\def\GetItemNUMBER#1{%
 \Point=\NumStack
 \advance\Point by #1
 \global\ItemNUMBER=\count\Point
}

% Note: \LeaveStack uses this routine. Do not destroy \Point
\def\GetItemTYPE#1{%
 \Point=\TypeStack
 \advance\Point by #1
 \global\ItemTYPE=\count\Point
}

% Note: \LeaveStack uses this routine. Do not destroy \Point
\def\GetItemSPAN#1{%
 \Point\SpanStack
 \advance\Point by #1
 \global\ItemSPAN=\count\Point
}

% Note: \LeaveStack uses this routine. Do not destroy \Point
\def\GetItemBOX#1{%
 \Point=\BoxStack
 \advance\Point by #1
 \global\setbox\ItemBOX=\vbox{\copy\Point}
 \global\ItemSIZE=\ht\ItemBOX
 \global\advance\ItemSIZE by \dp\ItemBOX
 \TEMPCOUNT=\ItemSIZE
 \divide\TEMPCOUNT by \Leading
 \divide\TEMPCOUNT by 65536
 \advance\TEMPCOUNT \@ne
 \ItemSIZE=\TEMPCOUNT pt
 \global\multiply\ItemSIZE by \Leading
}

% item joins stack

\def\JoinStack{%
 \ifnum\LengthOfStack=\MaxItems % stack is full of items
  \Warn{WARNING: Stack is full...some items will be lost!}
 \else
  \Point=\StatusStack
  \advance\Point by \LengthOfStack
  \global\count\Point=\ItemSTATUS
  \Point=\NumStack
  \advance\Point by \LengthOfStack
  \global\count\Point=\ItemNUMBER
  \Point=\TypeStack
  \advance\Point by \LengthOfStack
  \global\count\Point=\ItemTYPE
  \Point\SpanStack
  \advance\Point by \LengthOfStack
  \global\count\Point=\ItemSPAN
  \Point=\BoxStack
  \advance\Point by \LengthOfStack
  \global\setbox\Point=\vbox{\copy\ItemBOX}
  \global\advance\LengthOfStack \@ne
  \ifnum\ItemTYPE=\Figure % used in \MakePage
   \global\MoreFigurestrue
  \else
   \global\MoreTablestrue
  \fi
 \fi
}

% item leaves stack
% #1=physical position of the item to be removed

\def\LeaveStack#1{%
 {\Iteration=#1
 \loop
 \ifnum\Iteration<\LengthOfStack
  \advance\Iteration \@ne
  \GetItemSTATUS{\Iteration}
   \advance\Point by \m@ne
   \global\count\Point=\ItemSTATUS
  \GetItemNUMBER{\Iteration}
   \advance\Point by \m@ne
   \global\count\Point=\ItemNUMBER
  \GetItemTYPE{\Iteration}
   \advance\Point by \m@ne
   \global\count\Point=\ItemTYPE
  \GetItemSPAN{\Iteration}
   \advance\Point by \m@ne
   \global\count\Point=\ItemSPAN
  \GetItemBOX{\Iteration}
   \advance\Point by \m@ne
   \global\setbox\Point=\vbox{\copy\ItemBOX}
 \repeat}
 \global\advance\LengthOfStack by \m@ne
}

% clean stack
% This routine scans through the stack and removes anything
% that does not have STATUS=\InStack.

\newif\ifStackNotClean

\def\CleanStack{%
 \StackNotCleantrue
 {\Iteration=\z@
  \loop
   \ifStackNotClean
    \GetItemSTATUS{\Iteration}
    \ifnum\ItemSTATUS=\InStack
     \advance\Iteration \@ne
     \else
      \LeaveStack{\Iteration}
    \fi
   \ifnum\LengthOfStack<\Iteration
    \StackNotCleanfalse
   \fi
 \repeat}
}

% Find item.
% This macro searches from the top to the bottom of the
% stack for an item of a specified type and number.
% #1=type, #2=number
% If the specified item is found, then \StackPointer is set
% to point to it, else \StackPointer=-1.
% This routine is used to find the next figure or table
% by number.

\def\FindItem#1#2{%
 \global\StackPointer=\m@ne % assume item isn't in stack for now
 {\Iteration=\z@
  \loop
  \ifnum\Iteration<\LengthOfStack
   \GetItemSTATUS{\Iteration}
   \ifnum\ItemSTATUS=\InStack
    \GetItemTYPE{\Iteration}
    \ifnum\ItemTYPE=#1
     \GetItemNUMBER{\Iteration}
     \ifnum\ItemNUMBER=#2
      \global\StackPointer=\Iteration
      \Iteration=\LengthOfStack % terminate loop
     \fi
    \fi
   \fi
  \advance\Iteration \@ne
 \repeat}
}

% Find next type
% This macro searches from the top to the bottom of the stack
% looking for the first item which has STATUS=\InStack.
% If it is a figure then a figure is what will be considered
% next by \MakePage else table.

\def\FindNext{%
 \global\StackPointer=\m@ne % assume stack is empty for now
 {\Iteration=\z@
  \loop
  \ifnum\Iteration<\LengthOfStack
   \GetItemSTATUS{\Iteration}
   \ifnum\ItemSTATUS=\InStack
    \GetItemTYPE{\Iteration}
   \ifnum\ItemTYPE=\Figure
    \ifMoreFigures
      \global\NextItem=\Figure
      \global\StackPointer=\Iteration
      \Iteration=\LengthOfStack % terminate loop
    \fi
   \fi
   \ifnum\ItemTYPE=\Table
    \ifMoreTables
      \global\NextItem=\Table
      \global\StackPointer=\Iteration
      \Iteration=\LengthOfStack % terminate loop
    \fi
   \fi
  \fi
  \advance\Iteration \@ne
 \repeat}
}

% Change status
% Macro to change the status of a specified item in stack.
% #1=item, #2=new status

\def\ChangeStatus#1#2{%
 \Point=\StatusStack
 \advance\Point by #1
 \global\count\Point=#2
}

%--------------------------------------------------------%
%                       MAKEPAGE                         %
%--------------------------------------------------------%

% This macro is called at the start of every new page
% including the first. It scans through the stack picking
% out items which should be placed on this page. It then
% leaves space for the items to be placed later. The routine
% terminates when either there is no room on the page to
% fit the next figure or table, or there are no more items
% in the stack.

\def\Zone{\InZoneA}

\ZoneBAdjust=\z@

\def\MakePage{% allocate space on this page for stack items
 \global\ZoneBSize=\PageHeight
 \global\TextSize=\ZoneBSize
 \global\ZoneAFullPagefalse
 \global\topskip=\TextLeading
 \MakePageInCompletetrue
 \MoreFigurestrue
 \MoreTablestrue
 \FigInZoneBfalse
 \FigInZoneCfalse
 \TabInZoneBfalse
 \TabInZoneCfalse
 \global\FirstSingleItemtrue
 \global\FirstZoneAtrue
 \global\setbox\ZoneABOX=\box\VOIDBOX
 \global\setbox\ZoneBBOX=\box\VOIDBOX
 \global\setbox\ZoneCBOX=\box\VOIDBOX
 \loop
  \ifMakePageInComplete
 \FindNext
 \ifnum\StackPointer=\m@ne
  \NextItem=\m@ne
  \MoreFiguresfalse
  \MoreTablesfalse
 \fi
 \ifnum\NextItem=\Figure
   \FindItem{\Figure}{\NextFigure}
   \ifnum\StackPointer=\m@ne \global\MoreFiguresfalse
   \else
    \GetItemSPAN{\StackPointer}
    \ifnum\ItemSPAN=\Single \def\Zone{\InZoneB}\relax
     \ifFigInZoneC \global\MoreFiguresfalse\fi
    \else
     \def\Zone{\InZoneA}
     \ifFigInZoneB \def\Zone{\InZoneC}\fi
    \fi
   \fi
   \ifMoreFigures\Print{}\FigureItems\fi
 \fi
\ifnum\NextItem=\Table
   \FindItem{\Table}{\NextTable}
   \ifnum\StackPointer=\m@ne \global\MoreTablesfalse
   \else
    \GetItemSPAN{\StackPointer}
    \ifnum\ItemSPAN=\Single\relax
     \ifTabInZoneC \global\MoreTablesfalse\fi
    \else
     \def\Zone{\InZoneA}
     \ifTabInZoneB \def\Zone{\InZoneC}\fi
    \fi
   \fi
   \ifMoreTables\Print{}\TableItems\fi
 \fi
   \MakePageInCompletefalse % assume page is complete
   \ifMoreFigures\MakePageInCompletetrue\fi
   \ifMoreTables\MakePageInCompletetrue\fi
 \repeat
%\Print{TextSize=\the\TextSize}
%\Print{ZoneBSize=\the\ZoneBSize}
 \ifZoneAFullPage
  \global\TextSize=\z@
  \global\ZoneBSize=\z@
  \global\vsize=\z@\relax
  \global\topskip=\z@\relax
  \vbox to \z@{\vss}
  \eject
 \else
 \global\advance\ZoneBSize by -\ZoneBAdjust
 \global\vsize=\ZoneBSize
 \global\hsize=\ColumnWidth
 \global\ZoneBAdjust=\z@
 \ifdim\TextSize<23pt
 \Warn{}
 \Warn{* Making column fall short: TextSize=\the\TextSize *}
 \vskip-\lastskip\eject\fi
 \fi
}

\def\MakeRightCol{% allocate space for the right column of text
 \global\TextSize=\ZoneBSize
 \MakePageInCompletetrue
 \MoreFigurestrue
 \MoreTablestrue
 \global\FirstSingleItemtrue
 \global\setbox\ZoneBBOX=\box\VOIDBOX
 \def\Zone{\InZoneB}
 \loop
  \ifMakePageInComplete
 \FindNext
 \ifnum\StackPointer=\m@ne
  \NextItem=\m@ne
  \MoreFiguresfalse
  \MoreTablesfalse
 \fi
 \ifnum\NextItem=\Figure
   \FindItem{\Figure}{\NextFigure}
   \ifnum\StackPointer=\m@ne \MoreFiguresfalse
   \else
    \GetItemSPAN{\StackPointer}
    \ifnum\ItemSPAN=\Double\relax
     \MoreFiguresfalse\fi
   \fi
   \ifMoreFigures\Print{}\FigureItems\fi
 \fi
 \ifnum\NextItem=\Table
   \FindItem{\Table}{\NextTable}
   \ifnum\StackPointer=\m@ne \MoreTablesfalse
   \else
    \GetItemSPAN{\StackPointer}
    \ifnum\ItemSPAN=\Double\relax
     \MoreTablesfalse\fi
   \fi
   \ifMoreTables\Print{}\TableItems\fi
 \fi
   \MakePageInCompletefalse % assume page is complete
   \ifMoreFigures\MakePageInCompletetrue\fi
   \ifMoreTables\MakePageInCompletetrue\fi
 \repeat
 \ifZoneAFullPage
  \global\TextSize=\z@
  \global\ZoneBSize=\z@
  \global\vsize=\z@\relax
  \global\topskip=\z@\relax
  \vbox to \z@{\vss}
  \eject
 \else
 \global\vsize=\ZoneBSize
 \global\hsize=\ColumnWidth
 \ifdim\TextSize<23pt
 \Warn{}
 \Warn{* Making column fall short: TextSize=\the\TextSize *}
 \vskip-\lastskip\eject\fi
\fi
}

\def\FigureItems{% Stack pointer points to next figure
 \Print{Considering...}
 \ShowItem{\StackPointer}
 \GetItemBOX{\StackPointer} % auto calculates ItemSIZE
 \GetItemSPAN{\StackPointer}
  \CheckFitInZone % check to see if item fits
  \ifnum\ItemFits=\Yes
   \ifnum\ItemSPAN=\Single
     \ChangeStatus{\StackPointer}{\InZoneB} % flag to be output
     \global\FigInZoneBtrue
     \ifFirstSingleItem
      \hbox{}\vskip-\BodgeHeight
     \global\advance\ItemSIZE by \TextLeading
     \fi
     \unvbox\ItemBOX\ItemSep
     \global\FirstSingleItemfalse
     \global\advance\TextSize by -\ItemSIZE% allocate space
     \global\advance\TextSize by -\TextLeading
   \else
    \ifFirstZoneA
     \global\advance\ItemSIZE by \TextLeading
     \global\FirstZoneAfalse\fi
    \global\advance\TextSize by -\ItemSIZE
    \global\advance\TextSize by -\TextLeading
    \global\advance\ZoneBSize by -\ItemSIZE
    \global\advance\ZoneBSize by -\TextLeading
    \ifFigInZoneB\relax
     \else
     \ifdim\TextSize<3\TextLeading
     \global\ZoneAFullPagetrue
     \fi
    \fi
    \ChangeStatus{\StackPointer}{\Zone}
    \ifnum\Zone=\InZoneC \global\FigInZoneCtrue\fi
  \fi
   \Print{TextSize=\the\TextSize}
   \Print{ZoneBSize=\the\ZoneBSize}
  \global\advance\NextFigure \@ne
   \Print{This figure has been placed.}
  \else
   \Print{No space available for this figure...holding over.}
   \Print{}
   \global\MoreFiguresfalse
  \fi
}

\def\TableItems{% Stack pointer points to next table
 \Print{Considering...}
 \ShowItem{\StackPointer}
 \GetItemBOX{\StackPointer} % auto calculates ItemSIZE
 \GetItemSPAN{\StackPointer}
  \CheckFitInZone % check to see of item fits in Zone
  \ifnum\ItemFits=\Yes
   \ifnum\ItemSPAN=\Single
    \ChangeStatus{\StackPointer}{\InZoneB}
     \global\TabInZoneBtrue
     \ifFirstSingleItem
      \hbox{}\vskip-\BodgeHeight
     \global\advance\ItemSIZE by \TextLeading
     \fi
     \unvbox\ItemBOX\ItemSep
     \global\FirstSingleItemfalse
     \global\advance\TextSize by -\ItemSIZE
     \global\advance\TextSize by -\TextLeading
   \else
    \ifFirstZoneA
    \global\advance\ItemSIZE by \TextLeading
    \global\FirstZoneAfalse\fi
    \global\advance\TextSize by -\ItemSIZE
    \global\advance\TextSize by -\TextLeading
    \global\advance\ZoneBSize by -\ItemSIZE
    \global\advance\ZoneBSize by -\TextLeading
    \ifFigInZoneB\relax
     \else
     \ifdim\TextSize<3\TextLeading
     \global\ZoneAFullPagetrue
     \fi
    \fi
    \ChangeStatus{\StackPointer}{\Zone}
    \ifnum\Zone=\InZoneC \global\TabInZoneCtrue\fi
   \fi
%   \Print{TextSize=\the\TextSize}
%   \Print{ZoneBSize=\the\ZoneBSize}
  \global\advance\NextTable \@ne
   \Print{This table has been placed.}
  \else
  \Print{No space available for this table...holding over.}
   \Print{}
   \global\MoreTablesfalse
  \fi
}

% These macros check to see if an item of ItemSIZE will
% fit in a particular zone. If it will, then ItemFits
% will be set true else false.

\def\CheckFitInZone{%
{\advance\TextSize by -\ItemSIZE
 \advance\TextSize by -\TextLeading
 \ifFirstSingleItem
  \advance\TextSize by \TextLeading
 \fi
 \ifnum\Zone=\InZoneA\relax
  \else \advance\TextSize by -\ZoneBAdjust
 \fi
 \ifdim\TextSize<3\TextLeading \global\ItemFits=\No
 \else \global\ItemFits=\Yes\fi}
}

\def\BeginOpening{%
  % start 9pt a.s.a.p. so that \New.. commands get a chance to init.
  \ninepoint
  \thispagestyle{titlepage}%
  \global\setbox\ItemBOX=\vbox\bgroup%
    \hsize=\PageWidth%
    \hrule height \z@
    \ifsinglecol\vskip 6pt\fi % Bodge, to get same pos. as two-column!
}

\let\begintopmatter=\BeginOpening  %  alias for \BeginOpening

\def\EndOpening{%
  \One%  1 line fixed space below opening
  \egroup
  \ifsinglecol
    \box\ItemBOX%
    \vskip\TextLeading plus 2\TextLeading% var. space: min 1, max 3 lines
    \@noafterindent
  \else
    \ItemNUMBER=\z@%
    \ItemTYPE=\Figure
    \ItemSPAN=\Double
    \ItemSTATUS=\InStack
    \JoinStack
  \fi
}

% Figures

\newif\if@here  \@herefalse

\def\no@float{\global\@heretrue}
\let\nofloat=\relax % only enabled for one column

\def\beginfigure{%
  \@ifstar{\global\@dfloattrue \@bfigure}{\global\@dfloatfalse \@bfigure}%
}

\def\@bfigure#1{%
  \par
  \if@dfloat
    \ItemSPAN=\Double
    \TEMPDIMEN=\PageWidth
  \else
    \ItemSPAN=\Single
    \TEMPDIMEN=\ColumnWidth
  \fi
  \ifsinglecol
    \TEMPDIMEN=\PageWidth
  \else
    \ItemSTATUS=\InStack
    \ItemNUMBER=#1%
    \ItemTYPE=\Figure
  \fi
  \bgroup
    \hsize=\TEMPDIMEN
    \global\setbox\ItemBOX=\vbox\bgroup
      \eightpoint\nostb@ls{10pt}%
      \let\caption=\fig@caption
      \ifsinglecol \let\nofloat=\no@float\fi
}

\def\fig@caption#1{%
  \vskip 5.5pt plus 6pt%
  \bgroup % grouping and size change needed for plate pages
    \eightpoint\nostb@ls{10pt}%
    \setbox\TEMPBOX=\hbox{#1}%
    \ifdim\wd\TEMPBOX>\TEMPDIMEN
      \noindent \unhbox\TEMPBOX\par
    \else
      \hbox to \hsize{\hfil\unhbox\TEMPBOX\hfil}%
    \fi
  \egroup
}

\def\endfigure{%
  \par\egroup % end \vbox
  \egroup
  \ifsinglecol
    \if@here \midinsert\global\@herefalse\else \topinsert\fi
      \unvbox\ItemBOX
    \endinsert
  \else
    \JoinStack
    \Print{Processing source for figure \the\ItemNUMBER}%
  \fi
}

% Tables

\newbox\tab@cap@box
\def\tab@caption#1{\global\setbox\tab@cap@box=\hbox{#1\par}}

\newtoks\tab@txt@toks
\long\def\tab@txt#1{\global\tab@txt@toks={#1}\global\table@txttrue}

\newif\iftable@txt  \table@txtfalse
\newif\if@dfloat    \@dfloatfalse

\def\begintable{%
  \@ifstar{\global\@dfloattrue \@btable}{\global\@dfloatfalse \@btable}%
}

\def\@btable#1{%
  \par
  \if@dfloat
    \ItemSPAN=\Double
    \TEMPDIMEN=\PageWidth
  \else
    \ItemSPAN=\Single
    \TEMPDIMEN=\ColumnWidth
  \fi
  \ifsinglecol
    \TEMPDIMEN=\PageWidth
  \else
    \ItemSTATUS=\InStack
    \ItemNUMBER=#1%
    \ItemTYPE=\Table
  \fi
  \bgroup
    \eightpoint\nostb@ls{10pt}%
    \global\setbox\ItemBOX=\vbox\bgroup
      \let\caption=\tab@caption
      \let\tabletext=\tab@txt
      \ifsinglecol \let\nofloat=\no@float\fi
}

\def\endtable{%
  \par\egroup % end \vbox
  \egroup
  \setbox\TEMPBOX=\hbox to \TEMPDIMEN{%
    \eightpoint\nostb@ls{10pt}%
    \hss
    \vbox{%
      \hsize=\wd\ItemBOX
      \ifvoid\tab@cap@box
      \else
        \noindent\unhbox\tab@cap@box
        \vskip 5.5pt plus 6pt%
      \fi
      \box\ItemBOX
      \iftable@txt
        \vskip 10pt%
        \noindent\the\tab@txt@toks
        \global\table@txtfalse
      \fi
    }%
    \hss
  }%
  \ifsinglecol
    \if@here \midinsert\global\@herefalse\else \topinsert\fi
      \box\TEMPBOX
    \endinsert
  \else
    \global\setbox\ItemBOX=\box\TEMPBOX
    \JoinStack
    \Print{Processing source for table \the\ItemNUMBER}%
  \fi
}

\def\UnloadZoneA{%
\FirstZoneAtrue
 \Iteration=\z@
  \loop
   \ifnum\Iteration<\LengthOfStack
    \GetItemSTATUS{\Iteration}
    \ifnum\ItemSTATUS=\InZoneA
     \GetItemBOX{\Iteration}
     \ifFirstZoneA \vbox to \BodgeHeight{\vfil}%
     \FirstZoneAfalse\fi
     \unvbox\ItemBOX\ItemSep
     \LeaveStack{\Iteration}
     \else
     \advance\Iteration \@ne
   \fi
 \repeat
}

\def\UnloadZoneC{%
\Iteration=\z@
  \loop
   \ifnum\Iteration<\LengthOfStack
    \GetItemSTATUS{\Iteration}
    \ifnum\ItemSTATUS=\InZoneC
     \GetItemBOX{\Iteration}
     \ItemSep\unvbox\ItemBOX
     \LeaveStack{\Iteration}
     \else
     \advance\Iteration \@ne
   \fi
 \repeat
}

%--------------------------------------------------------%
%                         DIAGNOSTICS                    %
%--------------------------------------------------------%

\def\ShowItem#1{% Show details of on item entry in stack
  {\GetItemAll{#1}
  \Print{\the#1:
  {TYPE=\ifnum\ItemTYPE=\Figure Figure\else Table\fi}
  {NUMBER=\the\ItemNUMBER}
  {SPAN=\ifnum\ItemSPAN=\Single Single\else Double\fi}
  {SIZE=\the\ItemSIZE}}}
}

\def\ShowStack{% 
 \Print{}
 \Print{LengthOfStack = \the\LengthOfStack}
 \ifnum\LengthOfStack=\z@ \Print{Stack is empty}\fi
 \Iteration=\z@
 \loop
 \ifnum\Iteration<\LengthOfStack
  \ShowItem{\Iteration}
  \advance\Iteration \@ne
 \repeat
}

\def\B#1#2{%
\hbox{\vrule\kern-0.4pt\vbox to #2{%
\hrule width #1\vfill\hrule}\kern-0.4pt\vrule}
}

%-------------------------------------------------------%
%             SINGLE COLUMN OUTPUT ROUTINE              %
%-------------------------------------------------------%

\newif\ifsinglecol   \singlecolfalse

\def\onecolumn{%
  \global\output={\singlecoloutput}%
  \global\hsize=\PageWidth
  \global\vsize=\PageHeight
  \global\ColumnWidth=\hsize
  \global\TextLeading=12pt
  \global\Leading=12
  \global\singlecoltrue
  \global\let\onecolumn=\relax%         stop them using \onecolumn again
  \global\let\footnote=\sing@footnote%  enable footnotes
  \global\let\vfootnote=\sing@vfootnote
  \ninepoint % reset \baselineskip after leading change
  \message{(Single column)}%
}

\def\singlecoloutput{%
  \shipout\vbox{\PageHead\pagebody\PageFoot}%
  \advancepageno
  \ifplate@page
    \shipout\vbox{%
      \sp@pagetrue
      \def\sp@type{plate}%
      \global\plate@pagefalse
      \PageHead\vbox to \PageHeight{\unvbox\plt@box\vfil}\PageFoot%
    }%
    \message{[plate]}%
    \advancepageno
  \fi
  \ifnum\outputpenalty>-\@MM \else\dosupereject\fi%
}

\def\ItemSep{\vskip\ItemSepamount\relax}

\def\ItemSepbreak{\par\ifdim\lastskip<\ItemSepamount
  \removelastskip\penalty-200\ItemSep\fi%
}

% Modify plain's \endinsert so that the mn's spacing is used

\let\@@endinsert=\endinsert % save plain's original \endinsert

\def\endinsert{\egroup % finish the \vbox
  \if@mid \dimen@\ht\z@ \advance\dimen@\dp\z@ \advance\dimen@12\p@
    \advance\dimen@\pagetotal \advance\dimen@-\pageshrink
    \ifdim\dimen@>\pagegoal\@midfalse\p@gefalse\fi\fi
  \if@mid \ItemSep\box\z@\ItemSepbreak
  \else\insert\topins{\penalty100 % floating insertion
    \splittopskip\z@skip
    \splitmaxdepth\maxdimen \floatingpenalty\z@
    \ifp@ge \dimen@\dp\z@
    \vbox to\vsize{\unvbox\z@\kern-\dimen@}% depth is zero
    \else \box\z@\nobreak\ItemSep\fi}\fi\endgroup%
}

% Footnotes (only enabled in single column)

\def\gobbleone#1{}
\def\gobbletwo#1#2{}
\let\footnote=\gobbletwo % Gobble footnote's unless enabled by \onecolumn
\let\vfootnote=\gobbleone

\def\sing@footnote#1{\let\@sf\empty % parameter #2 (the text) is read later
  \ifhmode\edef\@sf{\spacefactor\the\spacefactor}\/\fi
  \hbox{$^{\hbox{\eightpoint #1}}$}\@sf\sing@vfootnote{#1}%
}

\def\sing@vfootnote#1{\insert\footins\bgroup\eightpoint\b@ls{9pt}%
  \interlinepenalty\interfootnotelinepenalty
  \splittopskip\ht\strutbox % top baseline for broken footnotes
  \splitmaxdepth\dp\strutbox \floatingpenalty\@MM
  \leftskip\z@skip \rightskip\z@skip \spaceskip\z@skip \xspaceskip\z@skip
  \noindent $^{\scriptstyle\hbox{#1}}$\hskip 4pt%
    \footstrut\futurelet\next\fo@t%
}

% Kill footnote rule
\def\footnoterule{\kern-3\p@ \hrule height \z@ \kern 3\p@}

\skip\footins=19.5pt plus 12pt minus 1pt
\count\footins=1000
\dimen\footins=\maxdimen

% Landscape

\def\landscape{%
  \global\TEMPDIMEN=\PageWidth
  \global\PageWidth=\PageHeight
  \global\PageHeight=\TEMPDIMEN
  \global\let\landscape=\relax%         stop them using \landscape again.
  \onecolumn
  \message{(landscape)}%
  \raggedbottom
}

%-------------------------------------------------------%
%               TWO COLUMN OUTPUT ROUTINE               %
%-------------------------------------------------------%

\output{%
  \ifLeftCOL
    \global\setbox\LeftBOX=\vbox to \ZoneBSize{\box255\unvbox\ZoneBBOX}%
    \global\LeftCOLfalse
    \MakeRightCol
  \else
    \setbox\RightBOX=\vbox to \ZoneBSize{\box255\unvbox\ZoneBBOX}%
    \setbox\MidBOX=\hbox{\box\LeftBOX\hskip\ColumnGap\box\RightBOX}%
    \setbox\PageBOX=\vbox to \PageHeight{%
      \UnloadZoneA\box\MidBOX\UnloadZoneC}%
    \shipout\vbox{\PageHead\box\PageBOX\PageFoot}%
    \advancepageno
    \ifplate@page
      \shipout\vbox{%
        \sp@pagetrue
        \def\sp@type{plate}%
        \global\plate@pagefalse
        \PageHead\vbox to \PageHeight{\unvbox\plt@box\vfil}\PageFoot%
      }%
      \message{[plate]}%
      \advancepageno
    \fi
    \global\LeftCOLtrue
    \CleanStack
    \MakePage
  \fi
}

% Startup message

\Warn{\start@mess}

\newif\ifCUPmtplainloaded % for use in documents
\ifprod@font
  \global\CUPmtplainloadedtrue
\fi

\def\mnmacrosloaded{} % so articles can see if a format file has been used.

\catcode `\@=12 % @ signs are non-letters

% \dump

% end of mn.tex
\fi

% If your system has the AMS fonts version 2.0 installed, MN.tex can be
% made to use them by uncommenting the line: %\AMStwofontstrue
%
% By doing this, you will be able to obtain upright Greek characters.
% e.g. \umu, \upi etc.  See the section on "Upright Greek characters" in
% this guide for further information.

\newif\ifAMStwofonts
%\AMStwofontstrue

\ifCUPmtplainloaded \else
  \NewTextAlphabet{textbfit} {cmbxti10} {}
  \NewTextAlphabet{textbfss} {cmssbx10} {}
  \NewMathAlphabet{mathbfit} {cmbxti10} {} % for math mode
  \NewMathAlphabet{mathbfss} {cmssbx10} {} %  "   "    "
  \ifAMStwofonts
    \NewSymbolFont{upmath} {eurm10}
    \NewSymbolFont{AMSa} {msam10}
    \NewMathSymbol{\upi}     {0}{upmath}{19}
    \NewMathSymbol{\umu}     {0}{upmath}{16}
    \NewMathSymbol{\upartial}{0}{upmath}{40}
    \NewMathSymbol{\leqslant}{3}{AMSa}{36}
    \NewMathSymbol{\geqslant}{3}{AMSa}{3E}

    \let\leq=\leqslant \let\le=\leqslant
    \let\geq=\geqslant \let\ge=\geqslant
  \else
    \def\umu{\mu}
    \def\upi{\pi}
    \def\upartial{\partial}
  \fi
\fi

% Marginal adjustments using \pageoffset maybe required when printing
% proofs on a Laserprinter (this is usually not needed).
% Syntax: \pageoffset{ +/- hor. offset}{ +/- vert. offset}
% e.g.    \pageoffset{-3pc}{-4pc}

\pageoffset{-2.5pc}{0pc}

\loadboldmathnames

%%%%%%%%%%%%%\Referee   %  uncomment this for referee mode (double spaced)

% \pagerange, \pubyear and \volume are defined at the Journals office and
% not by an author.

% \onecolumn        % enable one column mode
% \letters          % for `letters' articles
\pagerange{1--7}    % `letters' articles should use \pagerange{Ln--Ln}

%%%%%%%%%%%%%%%%%%%%%% MNRAS REFERENCE MACROS
% \refrule produces a horizontal rule used in references with identical
% authors.
\def\refrule{\hbox to 3pc{\leaders\hrule depth-2pt height 2.4pt\hfill}}
\def\ref{\hangindent=1truecm}

\def\aj{AJ}

\def\aea{A\&A}
\def\ada{Ann. d'Astroph.}

\def\apj{ApJ}
\def\apjl{ApJL}

\def\mnras{MNRAS}

%

%%%%%%%%%%%%%%%%%%%%%%%QUANTITA' FISICHE

\def\km{{\rm\,km}}

\def\kpc{{\rm\,kpc}}

%%%%%%%%%%%%%%%%%%%%%%%%%%%%DEFIZINIONI COME NELLA TESI%%%%%%%%%%%%%%%%
\def\devac{\hbox{$\rm{de Vaucouleurs}\,\,$}}
\def\arg{(\ra,\mr,\beta)}
\def\err{\Theta}
\def\etal{\hbox{${\rm et\; al.}\,\,$}}

\def\ie{I_{\rm e}}

\def\iss{I_{**}}
\def\ish{I_{\rm *D}}
\def\ass{A_{**}}
\def\ash{A_{\rm *D}}

\def\kms{\hbox{${\rm km\,s^{-1}}$}}
\def\lb{\hbox{$L_{\rm B}$}}
\def\cs{\hbox{$c_2^*$}}

\def\ms{M_{*}}
\def\mh{M_{\rm D}}

\def\ml{\Upsilon_*}
\def\mr{{\cal R}}
\def\msol{M_\odot}

\def\ra{r_{\rm a}}

\def\rch{r_{\rm D}}
\def\rcs{r_{\rm *}}
\def\re{R_{\rm e}}
\def\rhr{\rho _{\rm D}(r)}
\def\rsr{\rho _*(r)}
\def\roh{\rho _{\rm D\circ}}

\def\ross{R_{\rm ap}}
\def\rt{r_{\rm t}}

\def\sa{s_{\rm a}}
\def\san{s_{\rm a}^2}
\def\sigc{\sigma^2_\circ}
\def\sz{\sigma_\circ}
\def\sr{\sigma _{\rm r}}
\def\sp{\sigma _{\rm P}}
\def\spr{\sigma _{\rm P}(R)}
\def\s2p{\sigma ^2_{\rm P}}
\def\srad{\sigma ^2_{\rm r}}

\def\sav{\sigma ^2_{\rm ap}}

\def\sa2Is{\sigma ^{2I}_{\rm a*}}
\def\sigs{\Sigma _*}

%%%%%%%%%%%%%%%%%%%%%%%%%maggiore circa uguale
\catcode`\@=11
\def\gsim{\ifmmode{\mathrel{\mathpalette\@versim>}}
    \else{$\mathrel{\mathpalette\@versim>}$}\fi}
\def\lsim{\ifmmode{\mathrel{\mathpalette\@versim<}}
    \else{$\mathrel{\mathpalette\@versim<}$}\fi}
\def\@versim#1#2{\lower 2.9truept \vbox{\baselineskip 0pt \lineskip 
    0.5truept \ialign{$\m@th#1\hfil##\hfil$\crcr#2\crcr\sim\crcr}}}
\catcode`\@=12
%-------------
\let\umu=\mu \let\upi=\pi
% \font\euler=eurm10
% \def\umu{\hbox{$\euler\mu$}} \def\upi{\hbox{$\euler\pi$}}
% Uncomment the above two lines if the Euler font is available

%\Referee   %  uncomment this for referee mode

% \pagerange, \pubyear and \volume are defined at the Journals office and
% not by an author.

\pagerange{000}
\pubyear{1995}
\volume{000}
% \umufiche{}     % for articles with microfiche
% \authorcomment{}  % author comment for footline

\begintopmatter  %  start the two spanning material

\title{The Tilt of the Fundamental Plane of Elliptical Galaxies: I.
Exploring Dynamical and Structural Effects}

\author{L. Ciotti$^1$, B. Lanzoni$^2$, and A. Renzini$^{2,3}$}
\affiliation{$^1$Osservatorio Astronomico di Bologna, via Zamboni 33, 
    Bologna 40126, Italy}
\affiliation{}
\affiliation{$^2$Dipartimento di Astronomia, Universit\`a di Bologna,
     via Zamboni 33, Bologna 40126, Italy}
\affiliation{}
\affiliation{$^3$European Southern Observatory, Karl-Schwarzschild-Str. 2,
     Garching b. M\"unchen, Germany}
\shortauthor{Luca Ciotti, Barbara Lanzoni, Alvio Renzini}
\shorttitle{The Fundamental Plane of Galaxies}

% \acceptedline is to be defined at the Journals office and not
% by an author.

\acceptedline{}

\abstract 
{In this paper we explore several structural and dynamical effects on the 
projected velocity dispersion as possible causes of the fundamental plane (FP) 
tilt of elliptical galaxies. Specifically, we determine the size of the 
systematic trend along the FP in the orbital radial anisotropy, in the dark 
matter (DM) content and distribution relative to the bright matter, and in 
the shape of the light profile that would be needed to produce the tilt, 
under the assumption of a constant stellar mass to light ratio. Spherical, 
non rotating, two--components models are constructed, where the light 
profiles resemble the $R^{1/4}$ law. For the investigated models anisotropy
cannot play a major role in causing the tilt, while a systematic increase in 
the DM content and/or concentration may formally produce it. Also a suitable 
variation of the shape of the light profile can produce the desired effect, 
and there may be some observational hints supporting this possibility. 
However, fine tuning is always required in order to produce the tilt, while 
preserving the {\it tightness} of the galaxies distribution about the FP.}

\keywords {galaxies: elliptical and lenticular, cD -- dark matter -- structure}

\maketitle  %  finish the two spanning material

\section{Introduction}
Elliptical galaxies do not populate uniformly the three dimensional
parameter space having as coordinates the central velocity dispersion 
$\sz$, the effective radius $\re$, and the mean effective surface 
brightness $\ie=\lb /2\pi\re^2$, where $\lb$ is the total galaxy 
luminosity in the blue band. They rather closely cluster around a plane 
(Dressler \etal 1987, hereafter D87; Djorgovski \& Davis 1987;
Bender, Burstein \& Faber 1992, hereafter BBF; Djorgovski \& Santiago
1993; and references therein) thus called the Fundamental Plane (FP). 
The existence of these ``scaling relations" is believed to be of great 
importance for several reasons, including the understanding of the formation 
and evolution of elliptical galaxies, and their use as tracers of bulk motions, 
and potentially as a cosmological probe when studying the FP relations for 
clusters at higher and higher redshift. 

For their sample of Virgo ellipticals, BBF introduced a convenient 
coordinate system, where the axis are linear 
combinations of the observables $\log\sigc,~\log\re$ and $\log\ie$:
$$\eqalign{k_1 &\equiv (\log\sigc+\log \re)/\sqrt{2}\cr
                  k_2 &\equiv (\log\sigc+2\log \ie-\log \re)/\sqrt{6}\cr
                  k_3 &\equiv (\log\sigc-\log \ie-\log \re)/\sqrt{3},\cr}
\eqno(1)$$
and the FP is seen edge-on when projected on the $k_1-k_3$ plane. 
The FP for Virgo ellipticals is shown in Fig. 1 and follows the relation:
$$k_3=0.15\,k_1+0.36, \eqno(2)$$
having assumed a Virgo distance of 20.7 Mpc, and measuring $\sz$, $\re$,
and $\ie$ respectively in $\kms$, Kpc, and $L_{{\rm B}\odot}{\rm pc}^{-2}$ 
units (cfr. BBF). 
The two main properties of the FP for Virgo ellipticals are the so 
called {\it tilt}, i.e., the systematic increase of $k_3$ along the FP 
described by equation (2), and its tightness, i.e., the nearly 
constant and very small 
dispersion of $k_3$ at every location on the FP, with $\sigma(k_3)\simeq 0.05$. 

Using the virial theorem, the $k$-s can be related to the total galaxy mass 
$M=c_2\re\sigc$ by:
$$k_1={1\over\sqrt{2}}\log{M\over{c_2}},\eqno(3)$$
$$k_3={1\over\sqrt{3}}\log{2\pi M\over{\lb c_2}}. \eqno(4)$$
If the virial coefficient $c_2$ is constant for all the galaxies, the observed 
FP tilt as described by equation (2) implies a systematic 
trend in the mass to light ratio with galaxy luminosity: 
$M/\lb\propto \lb^{0.2}$ (e.g., D87). Meanwhile, 
the small and constant thickness of the distribution about the FP corresponds 
to a very small ($\lsim 12$ per cent) dispersion of $M/\lb$ for any given 
luminosity. The ``smallness" of the 0.2 exponent may give the impression of 
the FP tilt being just a minor, not especially important effect.
Yet, this is a misleading impression. As galaxies in Fig. 1 span a factor
$\sim 200$ in luminosity, the tilt corresponds to a factor 
$\sim 3$ increase of $M/\lb$ along the FP, from faint to bright galaxies. 

\beginfigure{1}
 \vskip 10cm
 \caption{{\bf Figure 1.} The distribution of Virgo (closed boxes) and Coma 
  (crosses) ellipticals in the ($k_1,k_2,k_3$) space, from BBF. The upper 
  panel shows the FP edge-on; in the lower panel the FP is seen nearly 
  face-on. Open circles represent our reference models $G_i$ (see Section 4.5).}
\endfigure

On the other hand, the virial coefficient $c_2$ depends on the mass, light and 
velocity dispersion distributions within the galaxy, and is 
constant only insofar all such distributions are homologous. It can be 
obtained by solving the Jeans equations, for any reasonable assumption
concerning the distribution of the bright and DM components, and the orbital
anisotropy, with the additional constraint that the star density distribution 
will reproduce the observed galaxy surface brightness profile. 

Considering only the stellar component (i.e., the fraction of the total mass 
whose density follows the light distribution), its mass $\ms$ can be expressed 
as:
$$\ms=\cs ~\re~\sigc,\eqno(5)$$
where $\cs=c_2$ in the case of a galaxy that is devoid of dark matter 
($\ms=M$), while and in general:
$${\ms\over\cs}={M\over{c_2}}.\eqno(6)$$
Substituting (6) in (3) and (4) we see that the origin of the FP tilt can be 
sought in two {\it orthogonal} directions:
either the tilt may arise from
a {\it stellar population} effect ($\ml\equiv\ms/\lb \propto \lb^{0.2}$ while 
$\cs$=const), or from a {\it structural/dynamical} effect ($\cs\propto 
\lb^{-0.2}$ 
while $\ml$=const). In the former case, the tilt would result from a trend 
in some combination of typical stellar metallicity, age, and initial mass 
function (IMF). A priori this appears to be a quite viable options: after all, 
a systematic trend in colours and line strengths is known to exists with 
galaxy luminosity (hence with $k_1$), which is usually ascribed to a trend in
the mean metallicity with the depth of the galactic potential well. However, 
the metallicity effect has been estimated to be marginal (D87; Djorgovski \& 
Santiago 1993), and indeed existing population synthesys models appear to 
fail to reproduce but a small fraction of the tilt, unless special conditions 
are verified (Renzini 1995). On the other hand, a drastic variation of the 
IMF along the FP is required to produce the tilt, with M/brown dwarfs turning 
from being a minor constituent to dominate the baryonic mass of ellipticals, 
yet with a very small dispersion in the IMF at any location on the FP 
(Renzini \& Ciotti 1993, hereafter RC). Searching for the origin of the FP 
tilt in this direction will be further pursued in a separate paper 
(Maraston, Renzini \& Ritossa 1996). 

In this paper we concentrate instead on the second option, assuming a constant 
stellar mass to light ratio $\ml$, and exploring under which conditions 
structural/dynamical effects may cause the tilt in $k_3$ via a systematic 
decrease of $\cs$. Although a contribution to the tilt may derive also 
from a trend in the rotational support (decreasing from faint to bright 
ellipticals; e.g., Davies \etal 1983), we concentrate here on the effects of 
systematic trends along the FP in 1) the degree of radial anisotropy of the 
velocity dispersion tensor, 2) the DM fraction and/or distribution within the 
galaxies, and 3) the density profile of the bright component. Concerning the 
dark matter, a preliminary exploration lead to conclude 
that the central regions of ellipticals should turn from baryon dominated to 
DM dominated with increasing $\lb$, again with fine tuning required to 
account for the tilt and yet preserve the observed small and constant thickness 
of the FP (RC). A break in the structural homology as a possible origin of 
the tilt has been suggested by Djorgovski (1995) and  Hjorth \& Madsen (1995). 

In Section 2 we briefly describe the dynamical models that
we have used for exploring points 1) and 2), and derive an analytical 
approximation for the virial coefficient $\cs$. In Section 3 we investigate 
anisotropy as a possible cause of the FP tilt, while in Section 4 we explore 
the effects of the amount and distribution of DM, in every case having 
assumed a fixed shape of the light profile for all the galaxies. Projected 
velocity dispersion profiles are computed for models that succeed in producing
the $k_3$ tilt.
In conjunction with available observations these profiles are then used to 
reject some classes of models and to suggest a future observational test for 
those that still survive. Having completed the analysis of the
{\it dynamical} options for the origin of the tilt, in Section 5 we pass
to investigate the {\it morphological} option. This is accomplished by
assuming isotropic models without DM, whose surface brightness distribution
is described by $R^{1/m}$ profiles, thus ascribing to a systematic variation
of $m$ the origin of the tilt. Finally, in Section 6 we discuss and summarize 
the results.

\section{The model galaxies}
We make use of four classes of two--components, spherical galaxy 
models, totally velocity-dispersion supported. We indicate with $r$ the 
spatial radial coordinate and with $R$ the projected one.

\subsection{Bright and dark matter distributions}
As well known, 
the empirical $R^{1/4}$ law (de Vaucouleurs 1948) suitably fits the observed 
surface brightness profiles of many elliptical galaxies, but its deprojection 
cannot be expressed in terms of elementary functions. For this reason we 
describe the stellar component by two density distributions which 
give a good approximation to the $R^{1/4}$ law when projected, and at the same 
time permit several fully analytical manipulations. We consider the Hernquist 
(1990) density law:
$$\rsr\,=\,{\ms\over{2\,\pi}}\,{\rcs\over{\,r\left(\rcs+r\right)^3}},
\eqno(7)$$
for which the effective radius $\re\,\simeq\,1.82\,\rcs$, and the Jaffe (1983) 
density law:
$$\rsr\,=\,{\ms\over{4\,\pi}}\,{\rcs\over{\,r^2\left(\rcs+r\right)^2}},
\eqno(8)$$
for which $\re\,\simeq\,0.76\,\rcs$.

The density law appropriate for dark haloes in elliptical galaxies is
yet to be determined. The same kind of density profile -- though with different 
masses and scale lengths, i.e., with $M_{\rm D}$ and $\rch$ replacing $\ms$ 
and $\rcs$, respectively -- may apply to describe the luminous and 
dark matter distributions in a scenario in which both are collisionless and 
have undergone similar dynamical processes during galaxy formation, yet starting
from different initial conditions (e.g., Bertin, Saglia \& Stiavelli 1992, 
hereafter BSS). 
On the other hand, the distributions of the dark and bright matter may have 
different shapes to the extent that the baryonic component has dissipated, thus 
sinking deeper into the potential well. 

In the first option, DM haloes present a central cusp (e.g., Dubinski \& 
Carlberg 1991; BSS; Kochanek 1993, 1994), and in this mood we investigate HH 
models, 
where both the luminous and the dark components are described by a Hernquist 
distribution, and JJ models where both follow a Jaffe profile. 
In the mood of dissipational collapse, we investigate two other classes 
of models 
where the dark halo density flattens at small radii, while 
the stellar distribution peaks towards the centre. In the first class, 
that we call HP models, the Hernquist 
luminous component is embedded in a Plummer (1911) dark halo:
$$\rhr\,=\,{3\,\mh\over{4\,\pi}}\,{\rch^2\over{\left(\rch^2+r^2
\right)^{5/2}}},\eqno(9)$$
in the other one (JQ models), the stellar component is described 
by the Jaffe formula and the dark halo by a truncated quasi--isothermal 
distribution:
$$\rhr =\cases{{\roh\,\rch^2(\rch^2+r^2)^{-1}}~~~&for $r\leq\rt$;\cr
~&~\cr
0~~~&for $r>\rt$.\cr}\eqno(10)$$

\subsection{The dynamical models}
In order to compare our models with the dynamical properties of the observed 
galaxies, we need their spatial and projected velocity dispersion 
profiles. 
These are obtained by solving the associated Jeans equation (see, e.g., 
Binney \& Tremaine 1987, hereafter BT):
$${d\,\rsr\,\srad(r)\over dr}+{2\alpha(r)\rsr\srad(r)\over r}=
\,-{G M(r)\over r^2}\,\rsr,\eqno(11)$$
with the boundary condition $\rsr\srad(r)\to 0\,~{\rm for}~r\to{\infty}$, 
where $M(r)$ is the mass within $r$.
We use for $\alpha(r)$ the Osipkov-Merritt formula:
$$\alpha(r)\equiv{1-{\sigma^2_{\theta}(r)\over{\srad(r)}}}=
{r^2\over{r^2+r{_a}^2}}\eqno(12)$$
(Osipkov 1979; Merritt 1985a,b),
so that the velocity dispersion tensor is nearly isotropic inside $\ra$ and 
radially anisotropic outside, consistently with N-body simulations (see, e.g., 
van Albada 1982). Analytical expressions for the radial velocity dispersion 
profiles of HH, HP and JJ models are given in the Appendix. 
Their projection is then obtained by (e.g., BT, p. 208):
$$\s2p(R)\,=\,{2\over\sigs(R)}\,\int_R^{\infty}{\left [{1-\alpha(r)}
{R^2\over{r^2}}\right ]}{{\rsr\sr^2(r)\,r}\over{\sqrt{r^2-R^2}}}dr,\eqno(13)$$
where $\sigs(R)=\ml\, I(R)$ is the surface stellar mass density.
The solution of (11) then provides the projected velocity 
dispersion profile, via equation (13).

Having fixed $\ml$ over all the FP, a galaxy model is therefore 
specified by 5 more parameters, namely $\ms$, $\rcs$ (or $\re$), $\mh$, 
$\rch$, and the anisotropy radius $\ra$. In the following, we replace $\mh$ and 
$\rch$ with the dimensionless ratios $\mr=\mh/\ms$ and $\beta=\rch/\rcs$.

The observed $\sz$ entering into 
the definitions of the $k$-s in (1) does not correspond to $\sp(0)$, but rather 
to the average over the aperture used for the spectrographic observations. 
That used by D87 for constructing the $k$-s of the Virgo galaxies in Fig. 1, 
was normalized to a $4\arcsec\times 4\arcsec$ aperture at Coma distance. 
Being the average of $\spr$ over such a rectangular aperture very similar to 
that over a circular aperture of $2\arcsec.2$ radius, we simulate $\sz$ by: 
$$\sav(\ross)=\,{2\,\pi\over M_{\rm P}^*(\ross)}{\int_0^{\ross}{\sigs(R)\,
    \s2p(R)\,R\,dR}},\eqno(14)$$ 
where $M_{\rm P}^*(\ross)$ is the projected stellar mass inside $\ross$.
We therefore mimics the actual observations considering $\sigma_{\rm ap}
\equiv\sz$, and we correspondingly get the virial coefficient $\cs$ from 
equation (5). The $k$-s are then obtained from (3), (4) and (6).

Instead of a fixed angular aperture (as used by D87), when 
calculating $\sigma_{\rm ap}$ we have for simplicity adopted a fixed 
{\it linear} aperture $\ross=0.02\,\re$ for all our models. As discussed in 
Section 4.3, such a choice has a negligible effect on our results. 

\subsection{An analytical approximation}
The described procedure to determine $\cs$ and the $k$-s for a specified 
set of parameters $\ra$, $\mr$, and $\beta$ was performed numerically solving 
(11)--(14) for a three--dimensional grid:
$\arg\in
[(\ra)_{\rm min},\infty[\times[0,10]\times[0.1,40]$, where $(\ra)_{\rm min}$ 
is the lower acceptable limit for the anisotropy radius as discussed in Section 
3. We express the resulting $\cs$ in the form:
$$\cs={A(\ross)\over{\err\arg}},\eqno(15)$$
where $A$ is the virial coefficient in absence of both DM and anisotropy, 
and $\err$ represents the correcting factor when 
such ingredients are included. In units of $\msol{\rm s}^{-2}\kpc^{-1}\km^{-2}$ 
the value of $A(0.02\re)$ is $1.74\,10^6$ and $6.49\,10^5$ for an Hernquist 
and a Jaffe stellar distribution, respectively. 
A good fit (within 5 per cent) for the numerical values of $\err$ for HH, 
HP and JJ models is given by:
$$\eqalign{\err(\ra,&\mr,\beta)={\left [{1}+{B\over{(\ra/\re) ^{\rm b}}}
\right ]}
\times\cr
&{\left\{1+\mr{C\over{\beta^{\rm c}
(\beta+D)^{\rm d}}}{\left [1+{E(\beta)\over{[(\ra/\re)+F(\beta)]^{{\rm f}
(\beta)}}}\right ]}\right\}}.\cr}\eqno(16)$$
with all the coefficients and exponents being reported in Tables 1 and 2.

\begintable{1}
\caption{{\bf Table 1.} Numerical values of the fitting $\beta$--independent 
parameters in equation (16).}
\halign{%
\rm#\hfil& \qquad\rm#\hfil& \qquad\rm\hfil#& \qquad\rm\hfil#& \qquad\rm\hfil#& 
\qquad\rm\hfil#& \qquad\rm\hfil#& \qquad\hfil\rm#\cr 
   & $B$ & b    & $C$  &  c   & $D$  &   d   \cr
\noalign{\vskip 10pt}
HH & 0.076 & 1.47 & 1.70 & 0.96 & 0.70 & 0.980 \cr
HP & 0.076 & 1.47 & 1.58 & 1.21 & 0.95 & 1.695 \cr
JJ & 0.009 & 1.36 & 1.06 & 0.48 & 0.05 & 0.515 \cr
}
\tabletext{}
\endtable

\begintable*{2}
\caption{{\bf Table 2.} Some numerical values of the fitting $\beta$--dependent 
parameters in equation (16) for HH (columns 2--4), HP (columns 5--7) and JJ 
models (columns 8--9; in this case, f$(\beta)=12.5$ for every value of $\beta$
).}
\halign{%
\rm#\hfil& \qquad\rm#\hfil& \qquad\rm#\hfil& \qquad\rm#\hfil& \qquad\rm#\hfil& 
\qquad\rm#\hfil& \qquad\rm#\hfil& \qquad\rm#\hfil& 
\qquad\rm#\hfil& \qquad\rm#\hfil& \qquad\rm\hfil#& \qquad\hfil\rm#\cr
 $~\beta$& &  $E(\beta)$& $F(\beta)$& f$(\beta)$& & $E(\beta)$& $F(\beta)$& 
f$(\beta)$& & $E(\beta)$& $F(\beta)$\cr
\noalign{\vskip 10 pt}

0.2& & -1.00   & 1.138& 4.40 & &   -1.00 & 1.278 & 4.10 & & -1.00   & 1.143 \cr
0.5& & -1.00   & 1.308& 4.70 & &   -1.00 & 1.191 & 6.00 & & -1.00   & 1.235 \cr
1.0& & \,\,1.00& 21.67& 4.50 & &\,\,1.00 & 0.946 & 3.55 & & -1.00   & 2.002 \cr
1.5& & \,\,1.00& 1.437& 4.60 & &\,\,2.82 & 1.115 & 3.73 & &\,\,1.00 & 1.312 \cr
2.0& & \,\,1.00& 1.278& 4.40 & &\,\,3.69 & 1.122 & 3.58 & &\,\,1.00 & 1.261 \cr
2.5& & \,\,1.00& 1.229& 3.90 & &\,\,3.90 & 1.086 & 3.39 & &\,\,1.00 & 1.238 \cr
3.0& & \,\,1.00& 1.176& 3.75 & &\,\,4.25 & 1.071 & 3.28 & &\,\,1.00 & 1.223 \cr
3.5& & \,\,1.00& 1.142& 3.55 & &\,\,4.44 & 1.051 & 3.16 & &\,\,1.00 & 1.214 \cr
4.0& & \,\,1.00& 1.111& 3.45 & &\,\,4.55 & 1.028 & 3.07 & &\,\,1.00 & 1.207 \cr
4.5& & \,\,1.00& 1.086& 3.35 & &\,\,4.57 & 1.004 & 2.97 & &\,\,1.00 & 1.202 \cr
5.0& & \,\,1.00& 1.063& 3.30 & &\,\,4.86 & 1.000 & 2.92 & &\,\,1.00 & 1.197 \cr
5.5& & \,\,1.00& 1.044& 3.25 & &\,\,5.08 & 0.996 & 2.88 & &\,\,1.00 & 1.194 \cr
6.0& & \,\,1.00& 1.028& 3.20 & &\,\,4.95 & 0.968 & 2.79 & &\,\,1.00 & 1.191 \cr
}
\endtable

The adopted form of the function $\err$ retains the main physical constraints 
of the problem. Indeed, $\err\to 1$ for $\ra\to\infty$, and $\mr\to 0$ or 
$\beta\to\infty$. 
The linear dependence on $\mr$ derives directly from (11).  
We did not attempt an analytic fit for the JQ models.

\subsection{Constraining models to the Fundamental Plane}
In order to constrain the models to lie on the FP reproducing its tilt, 
the value of $\err$ at each location on the FP is 
determined using (6), (4) and (15):
$$\err={A\over{2\,\pi}\,\ml}~{10^{0.26\,k_1+0.62}},\eqno(17)$$
and therefore the required trend in either $\ra$, $\mr$ or $\beta$ as a 
function of $k_1$ is derived. 
The stellar mass to light ratio $\ml$ is obtained from (17) with $k_1=2.6$ and 
$\err=1$, which corresponds to assume faintest galaxies to be isotropic and 
devoid of DM. For Hernquist models we find $\ml=5.5$, while for Jaffe ones 
we have $\ml=2.06$. 

\section{Making the FP tilt with a trend in the anisotropy}
In this section we ascribe the entire tilt of the FP to a trend with $\lb$ in 
the anisotropy degree of the galaxies (i.e., in $\ra$), assuming no DM. The 
values of the anisotropy radius at each location on the $k_1$ axis are 
determined by solving (17) and then (16) for $\ra$, with $\mr=0$. The results 
are shown in Fig. 2 for an Hernquist and a Jaffe stellar distribution. The 
curves are truncated because of the limits imposed by dynamical consistency: 
above a certain luminosity, in models constrained to the FP, the phase--space 
distribution function runs into negative values. The limits can easily be 
established in the frame of the Osipkov--Merritt relation for $\alpha(r)$, 
without having to know the distribution function of the system 
(Ciotti \& Pellegrini 1992). We find: $\ra\ge 0.25\rcs\simeq 0.138\re$ for the 
Hernquist models, and $\ra\gsim 0.05\rcs\simeq 0.036\re$ for the Jaffe ones. 

Thus, we conclude that anisotropy alone cannot be at the origin of the tilt, 
because the extreme values of $\ra$ that would be required correspond to 
dynamically inconsistent models. Note that another argument militate 
against radial anisotropy as the cause of the FP tilt: the requirement of 
radial orbit stability is much more stringent than the simple dynamical 
consistency, and $(\ra)_{\rm min}$ increases again. 

\beginfigure{2}
 \vskip 8cm
 \caption{{\bf Figure 2.} The trend of the anisotropy radius along the FP 
         required to produce its tilt, in Hernquist and Jaffe models. The 
         curves are truncated at the radius below which the models become 
         dynamically inconsistent. The band within dotted lines marks the 
         boundaries within which $\ra$ can vary at each location on the $k_1$ 
         axis in accordance with the observed FP tightness.}
\endfigure

\section{Making the FP tilt with a trend in either the dark matter fraction or 
 distribution}
Following the negative results of the previous Section, we assume global 
isotropy and move to explore DM as potentially 
responsible for the tilt of the
FP. We first ascribe all the tilt to a trend in the dark to bright mass ratio
$\mr$ at constant $\beta$, and then to a trend in the relative dark and
bright distributions $\beta$ at constant $\mr$.

\subsection{Varying the amount of DM}
We set $\ra=\infty$, $\beta=$const, and for $2.6\le k_1 \le 4.4$ we determine
the value of $\mr$ that is required to place the models on the FP.
Then $\mr$ is obtained from equations (16) and (17) 
for HH, HP and JJ models, and numerically for JQ models. 
Obviously, the larger $\beta$, the larger the variations 
of $\mr$ that are required to produce the tilt. Values of $\beta\lsim 1$ may 
have a mere academic interest, although some evidences seem 
to exist in support of a dark halo more centrally concentrated than the bright 
component (Saglia, Bertin \& Stiavelli 1992, hereafter SBS). By analogy with 
spiral galaxies, haloes are generally considered diffuse ($\beta > 1$), though 
in some cases with significant amounts of DM inside the half-light radius (SBS).

Concerning HH and HP models, for $\beta \simeq 5$ exceedingly large values of 
$\mr$ are required to produce the FP tilt ($\mr\simeq 30-175$), thus 
we conclude that an increasing DM content from 
faint to bright galaxies may be at the origin of the observed tilt,
provided that $\beta <5$.

The same problem affects all the JQ models that we have considered, for every 
values of $\beta$ and $\rt$. 

As regards JJ models, $\mr$ never becomes 
larger than 10 (for instance $\mr\simeq 9.5$ at the bright end of the FP for 
$\beta=5$), thus every value of this parameter is acceptable, for every 
explored value of $\beta$. Fig. 3 (upper panels) show the results 
for HH and JJ models, in the cases $\beta=1,2,5$.

\subsection{Varying the relative concentration of dark and bright matter}
We now assume the dark to bright matter ratio $\mr$ to be constant among the 
reference sample of elliptical galaxies, and ask the relative concentration of 
the two components ($\beta$) to produce the observed tilt in $k_3$. Thus, the 
values of $\beta$ along the FP are derived for $\ra\to\infty$ and 
$\mr=1,5,9$. For every class of models we find that if $\mr=1$, the DM in 
brightest galaxies should be more centrally concentrated than the luminous 
component ($\beta <1$). Values of $\beta >1$ at every location on the FP always 
require a prevalence of DM with respect to bright matter ($\mr > 1$), apart 
from JQ models which are again completely unsatisfactory, their values of 
$\beta$ being unrealistically small for every choice of $\mr$ (we therefore 
reject this class of models). Fig. 3 (lower panels) shows the trend of $\beta$ 
along the FP for HH and JJ models. 

\beginfigure*{3}
 \vskip 13cm
 \caption{{\bf Figure 3.} The trend along the FP of the DM content 
         (upper panels) at constant $\beta$ and that of the DM concentration 
         (lower panels) at constant $\mr$, required to produce the 
         tilt, in HH and JJ models. The band within dotted lines marks the 
         boundaries within which $\mr$ and $\beta$ can vary at each location 
         on the $k_1$ axis in accordance with the observed FP tightness.}
\endfigure

\subsection{Aperture effect}
Having used a fixed angular aperture, D87 have sampled a larger fraction of 
the total light (or effective radius) in fainter/smaller galaxies compared to 
brighter/larger galaxies. In fact, the effective radii of the galaxies in 
Fig. 1 range from $\sim 0.5$ kpc up to $\sim 10$ kpc, and therefore 
a circular aperture of $2\arcsec.2$ (i.e., $\sim 220$ pc radius at the adopted 
Virgo distance), corresponds to a circular region of $\sim 0.44\,\re$ radius 
at the faint end of the FP, and of only $\sim 0.02\,\re$ radius in galaxies at 
the bright end. When calculating $\sigma_{\rm ap}$, we have instead adopted 
a fixed linear aperture of $\ross=0.02\re$ radius, thus correctly simulating 
only the observed $\sz$ of the brightest galaxies and underestimating 
the fraction of effective radius sampled in the spectroscopic observations of 
fainter galaxies. Thus the derived $\cs$, and the corresponding 
values of the parameter responsible for the FP tilt, are biased by such a 
choice. Indeed, $\mr$ ($\beta$) is set equal to 0 ($\infty$) at the 
faint end of the FP, and therefore also the lower limit of the 
driving parameter is not affected by the {\it aperture bias}. Thus, the 
derived variation range of the parameters is correct, the main effects 
concerning the intermediate values of $\mr$ and $\beta$, and the curves in 
Fig. 3 may be modified in their shape only in the range between the starting 
and the ending points. On the other hand, a constant aperture radius for 
all the models have permitted us to express the correcting factor $\err$ 
in a simple analytical form. 

\subsection{Constraints from the tightness of the FP}
The narrow and nearly constant thickness of the galaxies distribution about the
FP (in the $k_3$ direction) corresponds to a very small ($\lsim 12$ per cent) 
dispersion in the ratio of 
$M/\lb$ to the corresponding virial coefficient. If $M/\lb$ ratios and virial 
coefficients are not finely anticorrelated, this implies
indeed a very small dispersion, separately for both quantities, at any location
on the FP. In the frame of our basic assumption ($\ml=$const), this sets a
very severe  restriction on $\cs$, hence on $\err$:
$${\delta\err\over{\err}}\lsim 0.12,\eqno(18)$$
which translates into strong constraints on the range that each parameter 
can span at any location on the FP. These can be easily derived analytically 
from equation (16), for the three classes of models. For HH and JJ, the dotted 
lines in Fig. 3 represent the band within which galaxy to galaxy variations of 
the corresponding parameter are allowed, and yet are consistent with the 
restrictions imposed by the tightness of the FP, i.e., with inequality (18).

It is evident from these figures that, whatever the structural parameter
that is responsible for the tilt of the FP, and whatever the assumed mass 
distribution, dramatic fine tuning is required to produce the tilt, and yet 
preserve the tightness of the FP (RC). 

Note that also $\delta\ra/\ra$ should be very small at each location on the FP 
(see the dotted band in Fig. 2), thus once more arguing against such an origin 
of the tilt.  

\subsection{Constraints from the velocity dispersion profiles}
In the assumption of global isotropy, the observation 
of the radial trend of $\spr$ may hopefully give insight on the DM content and 
distribution within the galaxies (e.g., SBS; Bertin \etal 1994; 
Carollo \& Danziger 1994a,b; Carollo \etal 1995; and references 
therein). A comparison between theoretical and observed $\sp$ profiles may 
therefore check the reliability of our models and test whether the DM is 
responsible for the FP tilt (cfr. RC). In this frame, we have computed the 
$\sp$ profiles of six reference models for every class, 
and for all the explored combinations of $\mr$ and $\beta$. 
We have chosen six FP locations ($k_1,k_2$) within the portion of the FP 
actually occupied by Virgo ellipticals (see Fig. 1, open circles):
$$\eqalign{G_1 &=(2.6,4.2);~~G_2=(3.3,3.2);~~G_3=(3.3,4.2);\cr ~~G_4 &=(4.2,3.0)
;~~G_5=(4.2,3.6);~~G_6=(4.4,3.4),}$$
where $G=G(k_1,k_2)$. $G_1$ corresponds to the faintest model, $G_6$
to the brightest one, while models in the pairs $G_2,G_3$ and $G_4,G_5$ only 
differ for the effective radius (and then surface brightness). Their 
luminosities and effective radii are determined inverting (1), (2) and (15):
$$\eqalign{\lb &=2\pi\,10^{1.15\,k_1-0.62}\cr
\re &=10^{1.07\,k_1-0.41\,k_2-0.21},\cr}$$
and are reported in Tables 3 and 4. For HH, HP and JJ models, 
some representative profiles are shown in Fig. 4 (namely, those corresponding 
to the cases where $\mr$ varies at constant $\beta=2$, and $\beta$ varies at 
constant $\mr=5$). The values of $\mr$ and $\beta$ in each model $G_i$ are 
reported in Tables 3 and 4.  

\beginfigure*{4}
 \vskip 15cm
 \caption{{\bf Figure 4.} Projected velocity dispersion profiles of HH, HP 
         and JJ isotropic models with variable $\mr$ and constant $\beta=2$ 
         (left panels), and with variable $\beta$ and constant $\mr=5$ 
         (right panels). The solid line corresponds to the faintest reference 
         model ($G_1$), while dotted, dashed, long dashed, dot-dashed, and 
         dot-long dashed lines correspond to models $G_2,\, ...,\, G_6$, 
         respectively.}
\endfigure

\begintable*{3}
\caption{{\bf Table 3.} HH and HP models: values of $\lb$ and $\re$ for the six 
reference models $G_i$, and the corresponding values of $\mr$ for constant 
$\beta=2$, and of $\beta$ for constant $\mr=5$ (columns 3-4 for HH models and 
5-6 for HP).}
\halign{
\rm#\hfil& \qquad\rm#\hfil& \qquad\rm#\hfil& \qquad\rm#\hfil& \qquad\rm#\hfil& 
\qquad\rm#\hfil& \qquad\rm#\hfil& \qquad\rm#\hfil& \qquad\hfil\rm#\cr 
\noalign{\vskip 3 pt}

$ ~$& $~~~~\lb$& $~~\re$& & $~~~~\mr$& $~~~~\beta$&  & $~~~~\mr$& 
$\beta$~~~~\cr     
$ ~$& $[10^{10}\msol$]& $[\kpc]$& & $(\beta=2)$& $(\mr=5)$&  & 
       $(\beta=2)$& $(\mr=5)$\cr     
\noalign{\vskip 10 pt}
$G_1$& ~~~0.15&      ~0.49& &     ~~~0.0 & ~~~~$\infty$& & ~~~0.0& $\infty$~~~\cr
$G_2$& ~~~0.96&      ~3.41& &     ~~~1.57& ~~~3.94& & ~~~4.75&    ~2.04~~\cr
$G_3$& ~~~~~{\tt "}& ~1.33& &~~~~~{\tt "}&~~~~~{\tt "}& &~~~~~{\tt "}& {\tt "}~~~~\cr
$G_4$& ~~10.55&      14.91& &     ~~~4.85& ~~~2.06& & ~14.66& ~1.25~~\cr
$G_5$& ~~~~~{\tt "}& ~8.48& &~~~~~{\tt "}&~~~~~{\tt "}& &~~~~~{\tt "}& {\tt "}~~~~\cr
$G_5$& ~~17.95&      13.62& &     ~~~5.85& ~~~1.84& & ~17.68& ~1.14~~\cr
\noalign{\vskip 3 pt}
}
\endtable

\begintable{4}
\caption{{\bf Table 4.} JJ models: values of $\lb$ and $\re$ for the six 
reference models $G_i$, and the corresponding values of $\mr$ for constant 
$\beta=2$, and of $\beta$ for $\mr=5$.}
\halign{
\rm#\hfil& \qquad\rm#\hfil& \qquad\rm#\hfil& 
\qquad\rm#\hfil& \qquad\rm#\hfil& \qquad\rm#\hfil& \qquad\hfil\rm#\cr 
\noalign{\vskip 3 pt}

$ ~$& $~~~~\lb$& $~~\re$& & $~~~~\mr$& $~~~~\beta$\cr     
$ ~$& $[10^{10}~\msol]$& $[{\rm Kpc}]$& & $(\beta=2)$& $(\mr=5)$\cr     
\noalign{\vskip 10 pt}
$G_1$& ~~~0.15&     ~0.49& & ~~~0.0 &     ~~~~$\infty$\cr
$G_2$& ~~~0.96&     ~3.41& & ~~~1.02&     ~~10.7\cr
$G_3$& ~~~~{\tt "}& ~1.33& & ~~~~~{\tt "}&~~~~~{\tt "}\cr
$G_4$& ~~10.55&     14.91& & ~~~3.16&     ~~~3.31\cr
$G_5$& ~~~~{\tt "}& ~8.48& & ~~~~~{\tt "}&~~~~~{\tt "}\cr
$G_5$& ~~17.95&     13.62& & ~~~3.81&     ~~~2.72\cr
\noalign{\vskip 3 pt}
}
\endtable

HH models, especially those at the bright end of the FP, are characterized by 
a sizable central depression in their $\sp$, while 
the observed profiles tipically decrease monotonically with radius, at least 
for $R \gsim 0.2\,\re$ (Carollo \& Danziger 1994a,b). Therefore the 
models which better agree with observations are those in which the off--centre 
maximum of $\sp$ lies inside this radius, i.e., those where $\mr$ varies at 
$\beta\lsim 2$, and where $\beta$ varies at $\mr\lsim 5$. In the case of HP 
models, the $\sp$-profiles always present a prominent off--center 
maximum, and they have to be rejected. On the 
contrary, the velocity dispersion profiles of JJ models are monotonically 
decreasing with radius for every explored values of $\mr$ and $\beta$, and 
therefore are consistent with observations. To permit a more quantitative 
comparison between the models and the observed velocity dispersion profiles, 
in the JJ case we define a slope of $\sp(R)$ as:
$$\Delta\sigma(R)=1-{\sp(R)\over{\sp(0)}},\eqno(19)$$
where $\sp(0)$ is the maximum (i.e., the central) value of the projected 
velocity dispersion. Table 5 gives the values of $\Delta\sigma$ 
for the six $G_{\rm i}$ models and for $R=2\,\re$, a typical value for
the outermost determinations of $\spr$ (Bertin \etal 1994). By these values 
we can recognize characteristic trends in the slope inside each scenario.

In fact, in the assumption that $\mr$ varies from faint to bright 
galaxies at $\beta$ constant less than unity, 
$\Delta\sigma(2\re)$ monotonically increases along the FP, i.e., the 
$\sp$-profiles systematically become steeper as galaxy luminosity increases. 
If $\beta=1$, instead, the slope of the $\sp$-profiles is the same for every 
model, no matter what is the luminosity, while if $\beta > 1$, it 
systematically decreases along the FP. This can be easily understood as 
the velocity dispersion profiles 
reflect the potential well of the systems. Thus, if $\beta <1$ an increasing 
$\mr$ along the FP corresponds to an increase of DM content in the inner 
regions of galaxies, while if $\beta >1$ external regions are involved. In 
the first case velocity dispersion increases at small radii and then $\sigma_P$ 
profiles steepen, while in the second one, effects 
concern external parts of profiles and thus they flatten along the FP. If 
$\beta =1$ instead, the potential well only deepens, but does not become 
narrower nor wider and velocity dispersion profiles do not vary their slope. 

In the frame of the second scenario ($\mr$ constant and $\beta$ decreasing 
along the FP) 
velocity dispersion profiles initially flatten and then 
steepen as galaxies luminosity increases. From previous assumptions $G_1$ is 
an isotropic and DM lacking galaxy, so the required condition 
$\beta (G_2)<\beta (G_1)$ is equivalent to have added a DM component in the 
external regions of the $G_2$ model. Consequentely the external velocity 
dispersion increases and $\spr$ flattens. Moving towards the brigth end 
of the FP, $\beta$ decreases and DM is more and more pushed towards central 
regions and there is a critical value $\beta_{\rm crit}$ when DM starts 
affecting the central parts of the $\spr$ profiles, rather than 
their external wings. Thus $\spr$ steepens for $\beta <\beta_{\rm crit}$.

\begintable{5}
\caption{{\bf Table 5.} Per cent slope of projected velocity dispersion 
profile as defined in equation (19) for JJ models.}
\halign{%
\rm#\hfil& \qquad\rm#\hfil& \qquad\rm\hfil#& \qquad\rm\hfil#& \qquad\hfil\rm#\cr
                & $G_1$& $G_2,G_3$& $G_4,G_5$& $G_6$\cr
\noalign{\vskip 10 pt}
$\mr(\beta=0.2)$& 57\% &  62\% &  67\% &  68\% \cr
\noalign{\vskip 1 pt  \vskip 1 pt}
$\mr(\beta=0.5)$& 57\% &  59\% &  61\% &  62\% \cr
\noalign{\vskip 1 pt  \vskip 1 pt}
$\mr(\beta=1.0)$& 57\% &  57\% &  57\% &  57\% \cr
\noalign{\vskip 1 pt  \vskip 1 pt}
$\mr(\beta=2.0)$& 57\% &  53\% &  50\% &  50\% \cr
\noalign{\vskip 1 pt  \vskip 1 pt}
$\mr(\beta=5.0)$& 57\% &  48\% &  43\% &  42\% \cr
\noalign{\vskip 1 pt  \vskip 1 pt}
$\beta(\mr=1.0)$& 57\% &  53\% &  61\% &  63\% \cr
\noalign{\vskip 1 pt  \vskip 1 pt}
$\beta(\mr=5.0)$& 57\% &  46\% &  46\% &  47\% \cr
\noalign{\vskip 1 pt  \vskip 1 pt}
$\beta(\mr=9.0)$& 57\% &  45\% &  42\% &  42\% \cr
}
\endtable

In conclusion, observations may in principle check whether the FP tilt can be 
ascribed to a trend of $\mr$ at $\beta=$const, or it is caused 
by a variation of $\beta$ at $\mr=$const, or whether a dynamical origin has 
to be rejected. In the first case it is also possible to 
determine whether DM in galaxies is more or less concentrated than the bright 
component, or if they are distributed in the same way. 

\section {Making the FP tilt with a trend in the surface brightness profile}
Systematic deviations of the ellipticals light distributions 
from the standard $R^{1/4}$ profile may also possibly cause the 
FP tilt (Djorgovski 1995; Hjorth \& Madsen 1995). 
In this Section we explore such a possibility through a class of models in 
which the log of the surface brightness is proportional to $R^{1/m}$ and 
$m$ is allowed to vary with galaxies luminosities. 
We assume global isotropy and no DM, thus $\cs$ depends only 
on the stellar density distribution, and we determine which
variation of $m$ along the FP is required to generate the tilt.

\subsection{The models} 
The surface brightness distribution of $R^{1/m}$ models 
is described by the generalized \devac law (Sersic 1968):
$$I(R)=I_\circ \exp[-b(m)(R/\re)^{1/m}],\eqno (20)$$
where $b(m)\simeq 2m-0.324$ for $0.5\leq m\leq 10$ (Ciotti 1991). 
The dynamical properties of this class of models are determined by solving 
equations (11)--(14) with $\alpha(r)=0$. In this case the virial coefficient 
$\cs$ depends on the aperture radius $\ross$ in a non trivial way. Thus, to 
correctly 
simulate real observations, equation (14) is solved by averaging $\spr$ over 
a fixed angular aperture of $1.\arcsec 6$ radius, i.e., a suitably varying 
$\ross/\re$ with galaxy luminosity. 

\subsection{The tilt and the tightness}
Assuming faintest galaxies to be $R^{1/4}$ systems, by analogy with Section 2, 
we set: 
$$\cs={A_4(\ross)\over{\err(m)}}, \eqno(21)$$
where $A_4$ is the virial coefficient for $m=4$ and $\err$ represents the 
correcting factor when $m\ne 4$. In order to produce the tilt, this has to 
increase from 1 to $\sim 3$ along the FP. The values of $m$ that force the six 
reference models $G_i$ to lie on the FP have been correspondingly determined:  
the result is that $m$ has to increase from 4 ($G_1$) up to $\sim 10$ ($G_6$) 
along the $k_1$, as shown in Fig. 5 (upper panel). The figure also shows the
band within which $m$ can vary for fixed luminosity consistently with the 
tightness of the FP. Once again, a fine tuning of the driving parameter is 
required to fit the observations: a very small ($\lsim 10$ per cent) scatter 
of $m$ at any location on the FP should be associated to a large variation of 
it with galaxies luminosities. 
If one assume instead $m=2$ for the faintest galaxies (model $G_1$), the 
required variation is even larger, about a factor 4, up to 
$\sim 8$ for the model $G_6$, and the permitted variation of it at each FP 
location remains very small (Fig. 5, lower panel).

\beginfigure{5}
 \vskip 10cm
 \caption{{\bf Figure 5.} The values of $m$ for the six reference models $G_i$ 
         required to produce the FP tilt in $R^{1/m}$ models, and the band 
         within which $m$ can vary at each location on the $k_1$ axis in 
         accordance with the observed FP tightness. In the upper panels, the 
         faintest model is characterized by $m=4$; in the lower panel, $m=2$ 
         at the faint end of the FP.}
\endfigure

\subsection{Comparison with observations}
It actually turns out that the surface brightness distribution of ellipticals 
is well described by $R^{1/m}$ profiles with variable $m$, any model 
with $3<m<10$ being hardly distinguishable from the $R^{1/4}$ law in the 
radial range usually covered by observations (Makino, Akiyama \& Sugimoto 
1990). However, a systematic trend of $m$ with galaxy luminosity has recentely 
been reported (Caon, Capaccioli \& D'Onofrio 1993, hereafter CCD), with $m$ 
increasing from $\sim 1$ up to $\sim 15$, thus spanning a much wider range 
than required to produce the tilt. Indeed, if one restricts to the Virgo 
galaxies in common with BBF but three (NGC 4406, NGC 4552 and NGC 4621), 
$m$ ranges between $\sim 2$ and $\sim 8$, in good agreement with the required 
increase of $m$ from faint to brigth galaxies. Worrysome is the 
apparently large dispersion inferred by observations, with $m$ varying by a 
factor $\sim 3$ at any given luminosity (see Fig. 6), at variance with 
the observed FP tightness. 

\beginfigure{6}
 \vskip 15cm
 \caption{{\bf Figure 6.} Values of $m$ as fitted by CCD along the major 
(upper panel), equivalent (middle panel) and minor (lower panel) axis light 
profile for the Virgo ellipticals in common with BBF. The solid line is the 
data points best fit line; the dotted lines mark the boundary of the 
permitted variation band of $m$, in accordance with the observed FP tightness 
(the same as in Fig. 5, lower panel).}
\endfigure

%Perhaps part of the discrepancy comes from observational difficulties, also 
%apparent from the very different values of $m$ often found along the minor 
%and the major axis light profile of the same object (even if classified as a 
%spherical galaxy). Some problem may also derive from the fitting procedure 
%used by CCD : in the proper generalized \devac law (20), $\re$ is 
%the half-light radius, while in their approach it is instead a mere
%fit parameter with no physical meaning. 
%In principle, a finely tuned anticorrelation of $\ml$ with $m$ could 
%compensate the extreme scatter of $m$ in such a way to maintain
%the FP tilt and thickness within the observed limits; yet this kind of 
%conspiracy appears very unlikely.

At variance with $m$ as possible cause of the FP tilt seems a conclusion 
that may be implicit in the BBF study. 
Being the ratio between tidal radius $\rt$ and core radius $r_{\rm c}$ about 
100-300 for giant ellipticals, when described by King (1966) models, one may 
suppose faintest galaxies have $\rt/r_{\rm c}=100$ and brightest ones 
$\rt/r_{\rm c}=300$, thus considering a trend in the bright matter distribution 
along the FP. However, as shown in BBF Fig. 5, the corresponding decrease in 
the value of $\cs$ is not sufficient to account for the FP tilt, in contrast 
with our result. However, King models are characterized by a flat core, 
at variance with high--resolution ground--based and HST observations 
which suggest an increasing density towards the very central 
regions of elliptical galaxies (Lauer \etal 1992a,b). Therefore we 
cannot consider our result injured by such an argument. 

We conclude that further observational studies are required in 
order to determine whether a progression of light-profile shapes along the FP 
really exists among cluster ellipticals.

\section{Discussion and Conclusions}

In this paper we have investigated possible structural or dynamical origins
for the observed tilt of the fundamental plane of elliptical galaxies,
considering in turn a systematic variation along the FP in the radial orbital 
anisotropy, in the dark matter content or distribution, and in the shape of 
the surface brightness profile. In doing so we have varied one such parameter 
at a time, while keeping the other three constant.

Our exploration indicates that all structural/dynamical solutions 
to the fundamental plane problem are rather unappealing, though some are more 
so than others. This comes from the strong {\it fine tuning} that is 
required, no matter whether the driving parameter is the anisotropy radius
($\ra$), the amount of dark matter ($\mr$), its distribution relative to the
bright matter ($\beta$), or the shape of the surface brightness distribution 
($m$). 

In addition to this, we have excluded a trend in the anisotropy as possible 
cause of the tilt because it leads to physically inconsistent models, and 
specific arguments also militate against global dark matter content. 
To produce the tilt, the dark to bright matter ratio $\mr$ should increase 
along the FP, from its faint to its bright end. This is just the opposite trend 
that one expects from galactic wind formation models (e.g., Arimoto \& Yoshii 
1987), the only ones so far that naturally account for the increase of
metallicity (as measured by either broad band colors or the Mg$_2$ index)
with the depth of the potential well (as approximatively measured by $\sz$). 
Here, the deeper the potential well, the less baryonic material is
expelled in a supernova driven wind, thus leading to lower final value of $\mr$.
Dissipationless merging models -- that do not account for the metallicity-$\sz$
correlation -- would predict $\mr$ to remain constant after a merging event,
or in case decrease slightly, as less bound, preferentially dark material
may escape from the system during the merging event.
In conclusion, we do not see any good reason why the dark matter fraction 
should systematically increase along the FP, and we actually have hints it 
may decrease somewhat. We are therefore inclined to exclude the parameter 
$\mr$ from being responsible for the tilt.

Rather more attractive is instead the possibility of a systematic decrease
of $\beta$ along the FP. Qualitatively, a trend of this kind is indeed 
expected if bright/baryonic matter has dissipated deeper into the 
potential well of small/faint galaxies compared to big/bright ones. Actually,
towards the faint end of the FP galaxies are characterized by a higher
surface brightness and stellar density, and lower effective radii: this 
suggests that in these galaxies the stellar component is more centrally 
concentrated relative to dark matter compared to galaxies at the other end of 
the FP (Guzman, Lucey \& Bower 1993). Thus, a systematic variation of $\beta$ 
appears more plausible than the previous two alternatives, though the
fine tuning problem remains.

Somewhat analogous is the case of a systematic trend in the shape of the stellar
distribution, for which there appears to be some observational support
(CCD). If the surface brightness profile of ellipticals is well
described by a generalized de Vaucouleurs law ($R^{1/m}$-law), then an 
increase of $m$ by a factor of $\sim 2-4$ along the FP is sufficient to
produce the tilt. The only embarrassment we see with this solution is, again, 
the required fine tuning.

There remains the possibility of an {\it hybrid} origin of the tilt, with more 
than one effect contributing to tilting the FP. For example, a small progression
of anisotropy, DM concentration, and shape ($m$), coupled with a stellar
population effect causing a modest increase of $\ml$. This is perhaps a 
reasonable solution of the tilt problem, yet a very difficult one to test
observationally, given that each effect may individually be buried in the 
observational noise, and yet the combination of all of them may conjure to
produce the observed tilt.

Whatever the solution, the tightness of the distribution of Virgo and Coma
ellipticals about the FP is clear evidence for a very standardized and
syncronized production of elliptical galaxies, at least those in clusters. 
Hypothetical formation 
processes that contain a great deal of stocasticity -- such as e.g., late
merging of spiral galaxies -- are likely to generate disparate final structures
(i.e., $\ra$, $\mr$, and $\beta$ distributions), and stellar age distributions,
hence large dispersions about the FP. Such scenarios are clearly disfavored by 
the very existence of a tight FP correlation.

This study has also shown that models where 
both dark and bright components follow a Jaffe density 
distribution (JJ models) may offer a better description of elliptical galaxies. 
In general, this applies to models where the DM distribution is similar to
the stellar one, albeit less concentrated (see also Dubinski \& Carlberg 1991; 
BSS; Kochanek 1993, 1994). Instead, centrally flat DM distributions
frequently give physically or astrophysically unacceptable results, such as 
velocity dispersion profiles that steeply increase outward, and negative 
values of their distribution function (Ciotti \& Pellegrini 1992). 
Furthermore, a core radius for the DM haloes is hardly justifiable for 
dissipationless formation, as the gravitational force is not characterized by 
any specific scale length. 

\section{Acknowledgments} 
We would like to thank Ralf Bender, James Binney, Nicola Caon, George 
Djorgovski, Mauro D'Onofrio, Silvia Pellegrini, and Massimo Stiavelli for 
useful discussions. This work was supported in part by the Italian 
Ministry of Research (MURST), and in part by the Consiglio Nazionale delle 
Ricerche (CNR).

%\vskip 30truecm
%\vfil\eject
\onecolumn
\appendix
\section{Velocity dispersion profiles for HH, HP, and JJ models}
With the assumed radial trend for the of the velocity dispersion
tensor, the solution of the Jeans equation can be written in integral
form, as shown by Binney (1980). Moreover, for our two-component
models the solutions can be obtained in closed form, and we give here
the resulting radial velocity dispersions. The tangential component can be
successively obtained by using eq. (12). In order to avoid a cumbersome
notation, we define the dimensionless variables $s\equiv r/\rcs$ and
$s_{\rm a}\equiv \ra/\rcs$. It is straightforward to show that the
general velocity dispersion for our models looks like:
$$\rsr\srad(r)=\,{{G\ms^2}\over{2\pi\rcs^4}}\,{
{(\ass+\san\,\iss)+\mr(\ash+\san\,\ish)}\over{s^2+\san}},
\eqno(A1)$$
where the functions $\ass$ and $\iss$ depend on the stellar density 
distribution only, and $\ish$ and $\ash$ are the interaction terms
due to the DM gravitational field. Note that when the velocity dispersion
is completely isotropic (i.e. $s_{\rm a}\to\infty$) only the $I$'s functions
remains; on the contrary, in the formal case of completely radial orbits
($s_{\rm a} =0$), the velocity dispersion is described by the $A$'s functions.
For the Hernquist and Jaffe stellar profiles, the two components due to the
pure stellar distribution are respectively given by:
$$\ass ={4s+1\over 12(1+s)^4},\eqno (A2)$$
$$\iss =-{12s^3+42s^2+52s+25\over 12(1+s)^4}-\ln{s\over 1+s},
\eqno (A3)$$
and
$$\ass =-{1\over 2}\left[{2s+3\over 2(1+s)^2}+\ln{s\over 1+s }\right],
\eqno (A4)$$
$$\iss =-{1\over 2}\left[
{(6s^2+6s-1)(2s+1)\over 2s^2(1+s)^2}+6\ln{s\over 1+s}\right].
\eqno (A5)$$
We give now the interactions terms for the various distributions.
Starting with the HH models, we have:
$$\ash =-{1\over 2(\beta-1)^2(1+s)^2}+
{\beta+1\over (\beta-1)^3(1+s)}+
{\beta\over (\beta-1)^3(\beta+s)}
+{2\beta+1\over (\beta-1)^4}\ln{1+s\over \beta+s},
\eqno (A6)$$
$$\ish=-{1\over 2(\beta-1)^2(1+s)^2}+
{1\over \beta(\beta-1)^3(\beta+s)}-
{\beta-3\over (\beta-1)^3(1+s)}-{\ln s\over \beta^2}+
{(\beta^2-4\beta+6)\ln (1+s)\over (\beta-1)^4}-
{(4\beta-1)\ln (\beta+s)\over \beta^2(\beta-1)^4}.
\eqno (A7)$$

~

~

~

For HP models the functions are more complicated, due to the different
qualitative behaviour of the density distributions. After lenghty
calculations, we find:
$$\ash ={5(2\beta^2-1)\over 2(1+\beta^2)^3}-
{s^3(10\beta^2-5)+s^2(2\beta^4+9\beta^2-8)+s(8\beta^4-9\beta^2-2)
+(5\beta^4-10\beta^2)\over 
2(1+\beta^2)^3(1+s)^2\sqrt{\beta^2+s^2}}-$$
$$\quad\quad\quad\quad
{2\beta^4-11\beta^2+2\over 2(1+\beta^2)^{7/2}}
{\ln{(1+s)(\sqrt{1+\beta^2}-1)\over
\sqrt{(1+\beta^2)(\beta^2+s^2)}+\beta^2-s}}
\eqno (A8)$$
and
$$\ish ={(2-13\beta^2)\over 2\beta^2(1+\beta^2)^3}+ 
{s^3(13\beta^2-2)+s^2(3\beta^4+14\beta^2-4)
+s(11\beta^4-6\beta^2-2)+(\beta^6+10\beta^4-6\beta^2)
\over 2(1+s)^2\beta^2\left(1+\beta^2\right)^3\sqrt{\beta^2+s^2}}+$$
$$\quad\quad\quad {3(\beta^2-4)\over
2(1+\beta^2)^{7/2}}{\ln {(1+s)(\sqrt{1+\beta^2}-1)\over
\sqrt{(1+\beta^2)(\beta^2+s^2)}+\beta^2-s}}.
\eqno (A9)$$
Finally, for JJ models, we have:
$$\ash =-{1\over 2}\left[{1\over (\beta-1)(s+1)}+{\ln s \over\beta}-
{(\beta-2)\ln (1+s)\over (\beta-1)^2}-
{\ln (\beta+s)\over \beta(\beta-1)^2}\right ],
\eqno (A10)$$
$$\ish =-{1\over 2}\left[{2s^2(3\beta^2-\beta-1)+s(3\beta^2-\beta-2)-
\beta(\beta-1)\over
2s^2 \beta^2(\beta-1)(1+s)}-
{(3\beta^2+2\beta+1)\ln s\over \beta^3 }+
{(3\beta-4)\ln (1+s)\over (\beta-1)^2}+
{\ln (\beta+s)\over \beta^3(\beta-1)^2}\right].
\eqno (A11)$$
Note that the singularity for $\beta=1$ in $\ash$ and $\ish$ for HH and JJ 
models is eliminable, for in the limit $\beta\to 1$ these expressions 
coincide with the corresponding $\ass$ and $\iss$. 

\section*{References}
\beginrefs
\bibitem Arimoto N., Yoshii Y., 1987,  \aea, 173, 23
\bibitem Bender R., Burstein D., Faber S.M., 1992, \apj, 399, 462 (BBF)
\bibitem Bertin G., Saglia R.P., Stiavelli M., 1992, \apj, 384, 423 (BSS)
\bibitem Bertin G., Bertola F., Buson L.M., Danziger I.J., Dejonghe H., 
         Sadler E.M., Saglia R.P., de Zeeuw P.T., Zeilinger W.W., 1994, 
         \aea, 292, 381
\bibitem Binney J., 1980, \mnras, 190, 873
\bibitem Binney J., Tremaine S., 1987, Galactic Dynamics (Princeton 
         University Press) (BT)
\bibitem Caon N., Capaccioli M., D'Onofrio M., 1993, \mnras, 265, 1013 (CCD)
\bibitem Carollo C.M., Danziger I.J., 1994a, \mnras, 270, 523
\bibitem Carollo C.M., Danziger I.J., 1994b, \mnras, 270, 743
\bibitem Carollo C.M., de Zeeuw P.T., van der Marel R.P., Danziger I.J., 
         Qian E.E., 1995, \apjl, 441, L25
\bibitem Ciotti L., 1991, \aea, 249, 99
\bibitem Ciotti L., Pellegrini S., 1992, \mnras, 255, 561
\bibitem Davies R.L., Efstathiou G., Fall S.M., Illingworth G., Schechter P.L., 
         1983, \apj, 266, 41
\bibitem de Vaucouleurs G., 1948, \ada, 11, 247
\bibitem Djorgovski S., 1995, \apj, 438, L29
\bibitem Djorgovski S., Davis M., 1987, \apj, 313, 59
\bibitem Djorgovski S., Santiago B.X., 1993, in
         Danziger I.J., Zeilinger W.W., Kj\"ar K., eds, 
         Structure, Dynamics and Chemical Evolution of Elliptical
         Galaxies. ESO, Garching, p. 59
\bibitem Dressler A., Lynden-Bell D., Burstein D., Davies R.L., Faber S.M.,
     Terlevich R.J., Wegner G., 1987, \apj, 313, 42 (D87)
\bibitem Dubinski J., Carlberg G., 1991, \apj, 378, 496
\bibitem Guzman, R., Lucey, J.R., Bower, R.G. 1993, \mnras, 265, 731
\bibitem Hernquist L., 1990, \apj, 356, 359
\bibitem Hjorth J., Madsen J., 1995, \apj, 445, 55
\bibitem Jaffe W., 1983, \mnras, 202, 995
\bibitem King I., 1966, Astron. J., 71, 64
\bibitem Kochanek C.S., \apj, 1993, 419, 12
\bibitem Kochanek C.S., \apj, 1994, 436, 56
\bibitem Lauer T.R., Faber S., Lynds C.R., Baum W.A., Ewald S.P., Groth E.J., 
         Hester J.J, Holtzman J.A., Light R.M., O\`Neil E.J., Schneider D.P., 
         Shaya E.J., Westphal J.A., 1992a, \aj, 103, 703
\bibitem Lauer T.R., Faber S., Currie D.G., Ewald S.P., Groth E.J., 
         Hester J.J, Holtzman J.A., Light R.M., O\`Neil E.J., Shaya E.J., 
         Westphal J.A., 1992b, \aj, 104, 552 
\bibitem Maraston C., Renzini, A., Ritossa C., 1996, in preparation
\bibitem Merritt D, 1985a, Astron. J., 90, 1027
\bibitem Merritt D, 1985b, \mnras, 214, 25P
\bibitem Osipkov L.P., 1979, Pis'ma Astr. Zh., 5, 77
\bibitem Plummer H.C., 1911, \mnras, 71, 460
\bibitem van Albada T.S., 1982, \mnras, 201, 939
\bibitem Renzini A., 1995, in Gilmore G., van der Kruit P., eds., 
         Stellar Populations, Kluwer, Dordrecht, p. 325
\bibitem Renzini A., Ciotti L., 1993, \apjl, 416, L4 (RC)
\bibitem Saglia R.P., Bertin G., Stiavelli M., 1992, \apj, 384, 433 (SBS)
\bibitem Sersic J.L., 1968, Atlas de galaxias australes, Observatorio 
         Astronomico, Cordoba
\endrefs
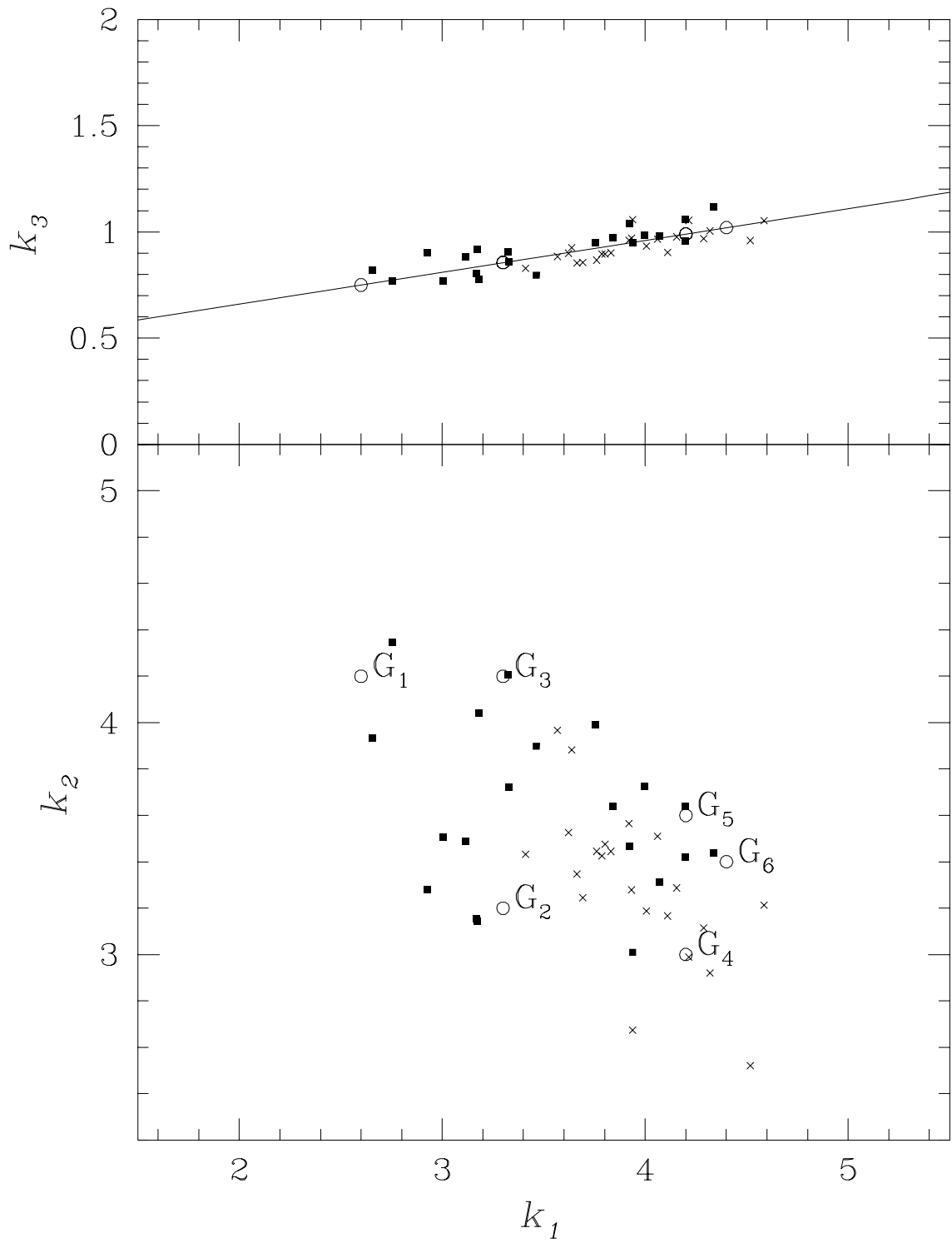
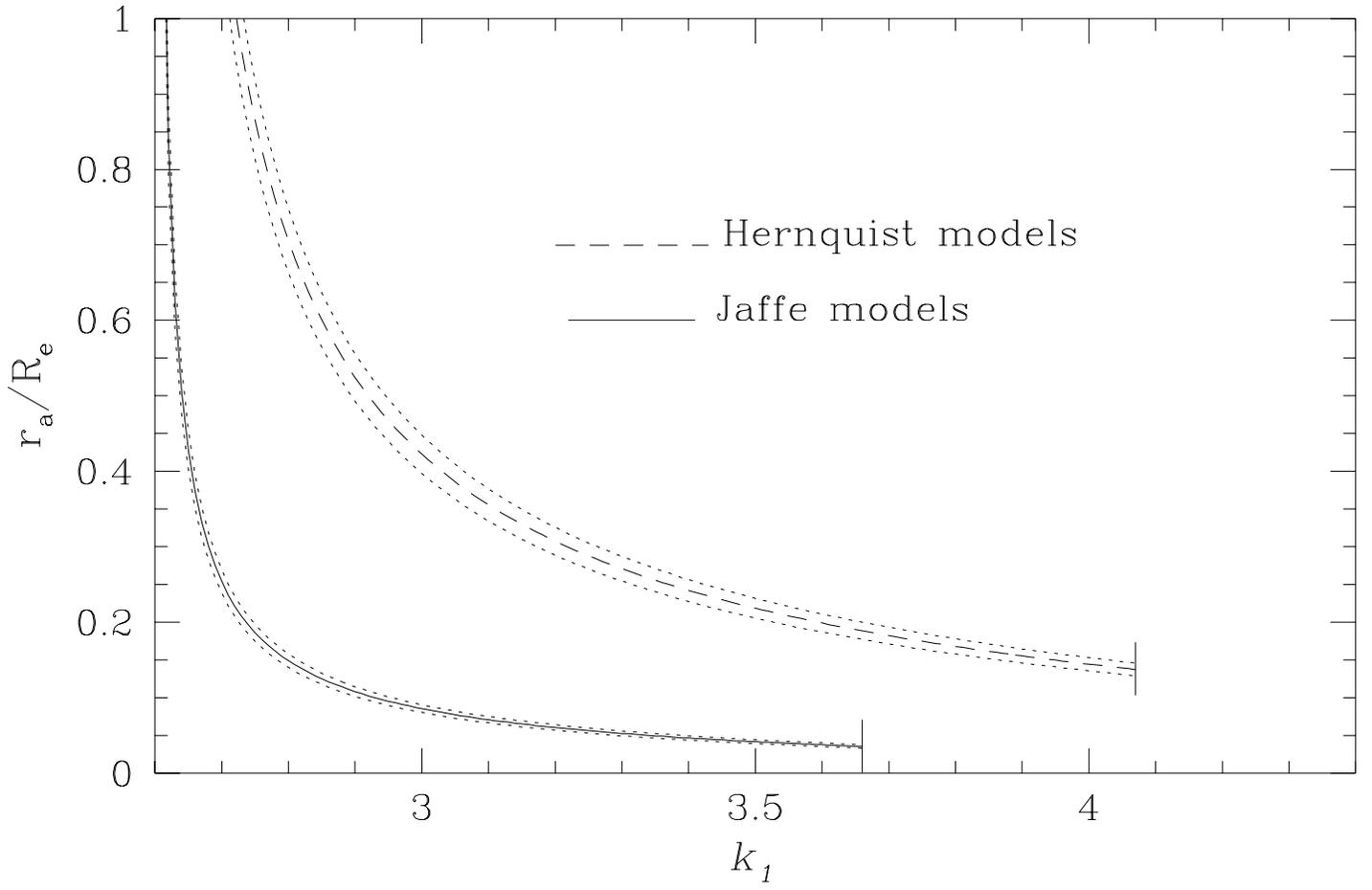
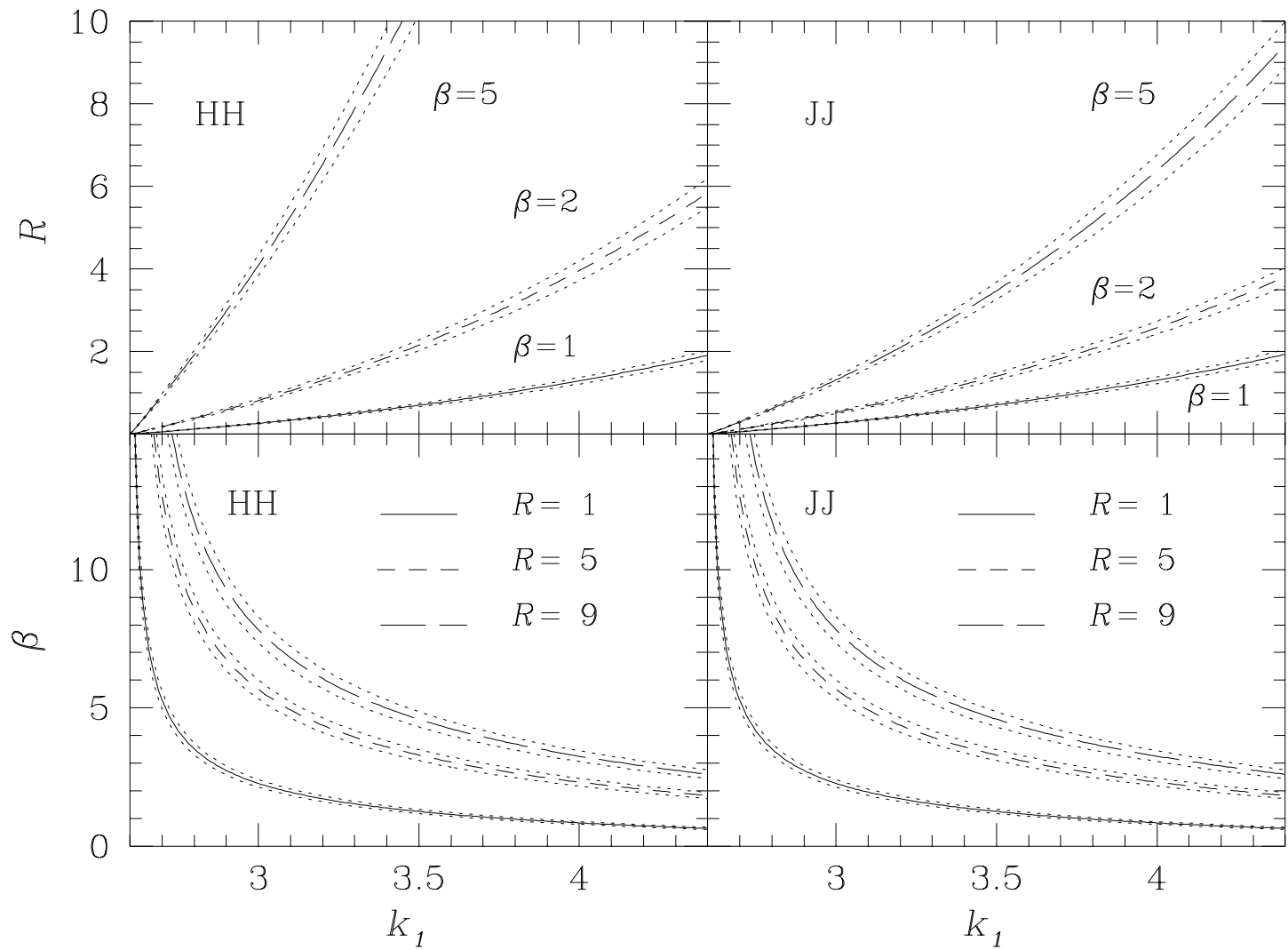
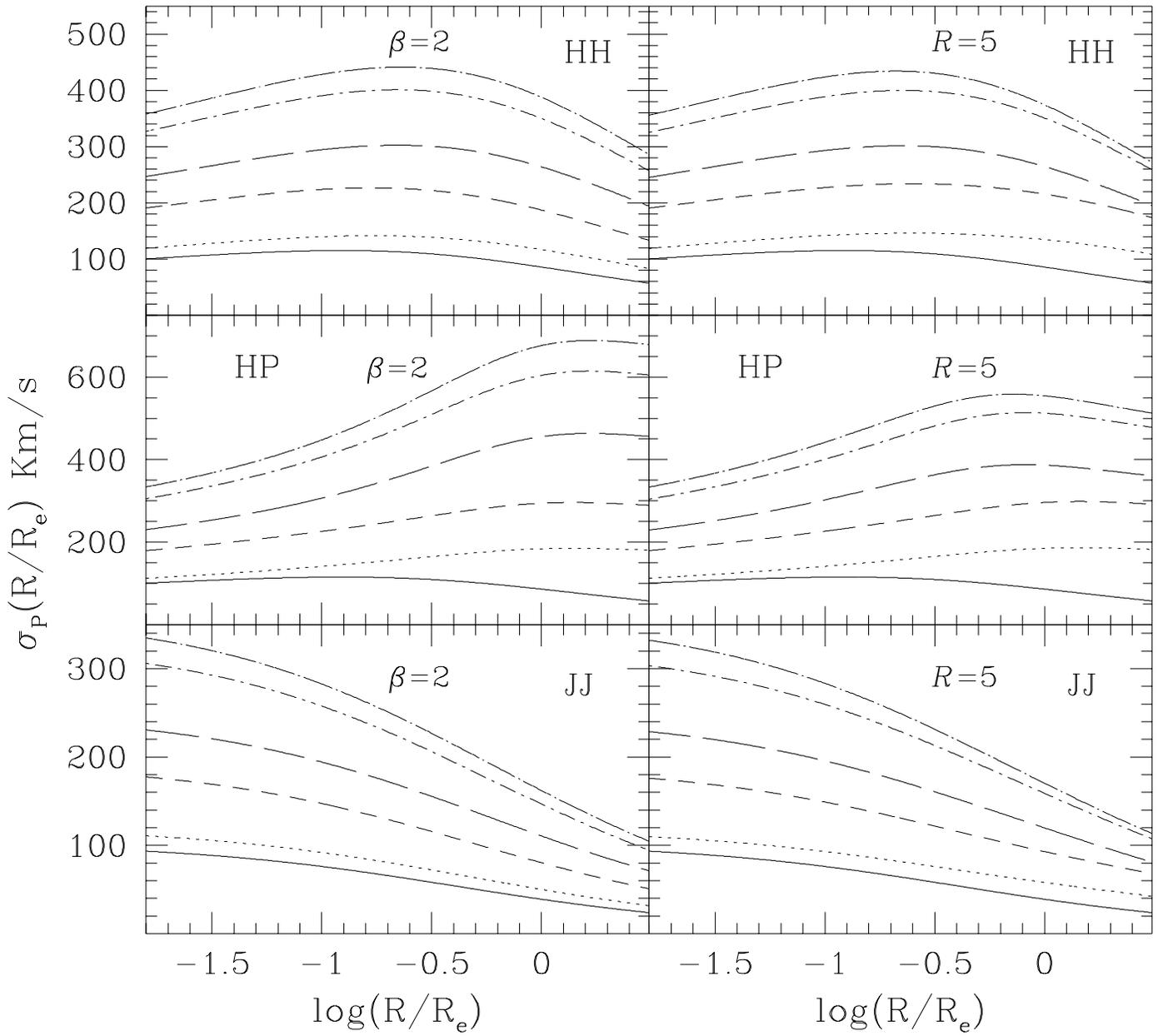
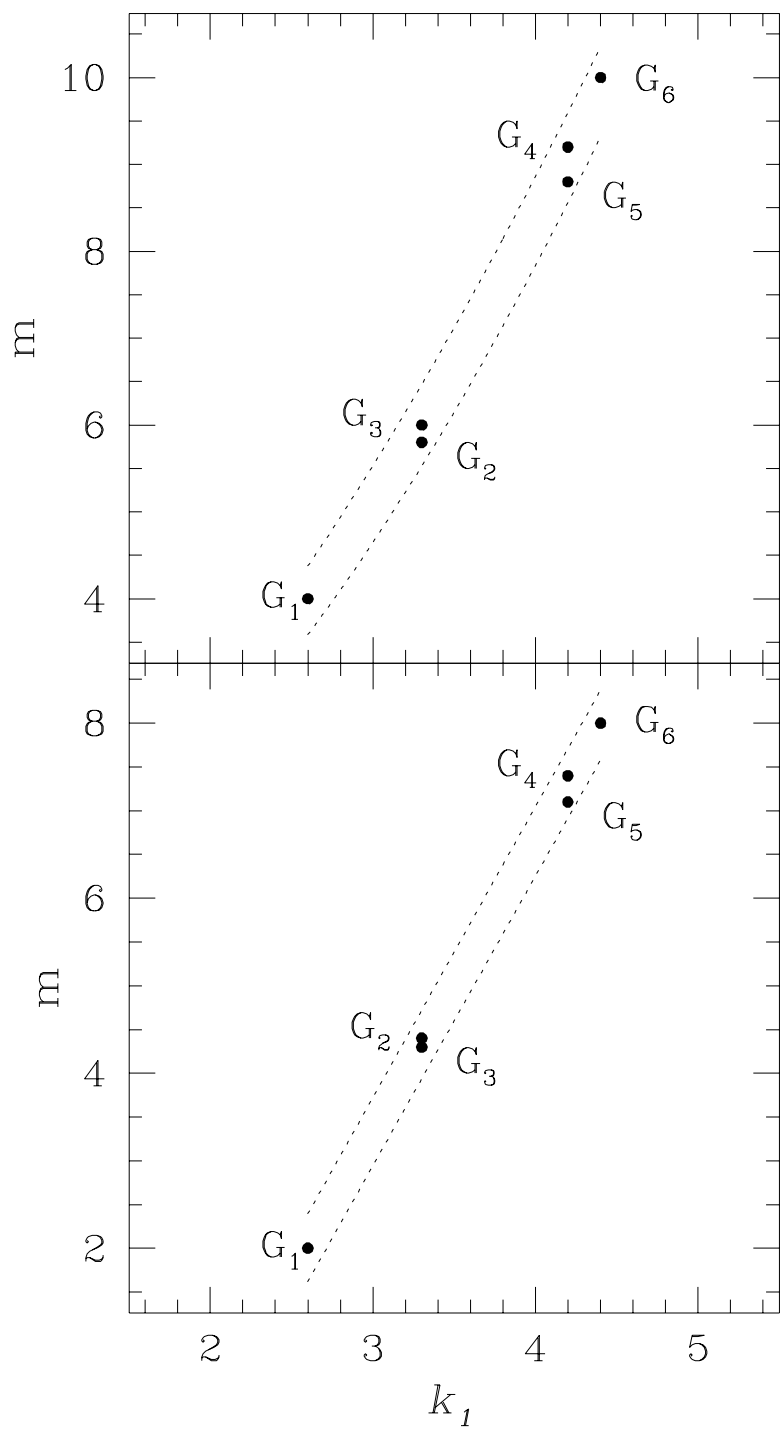
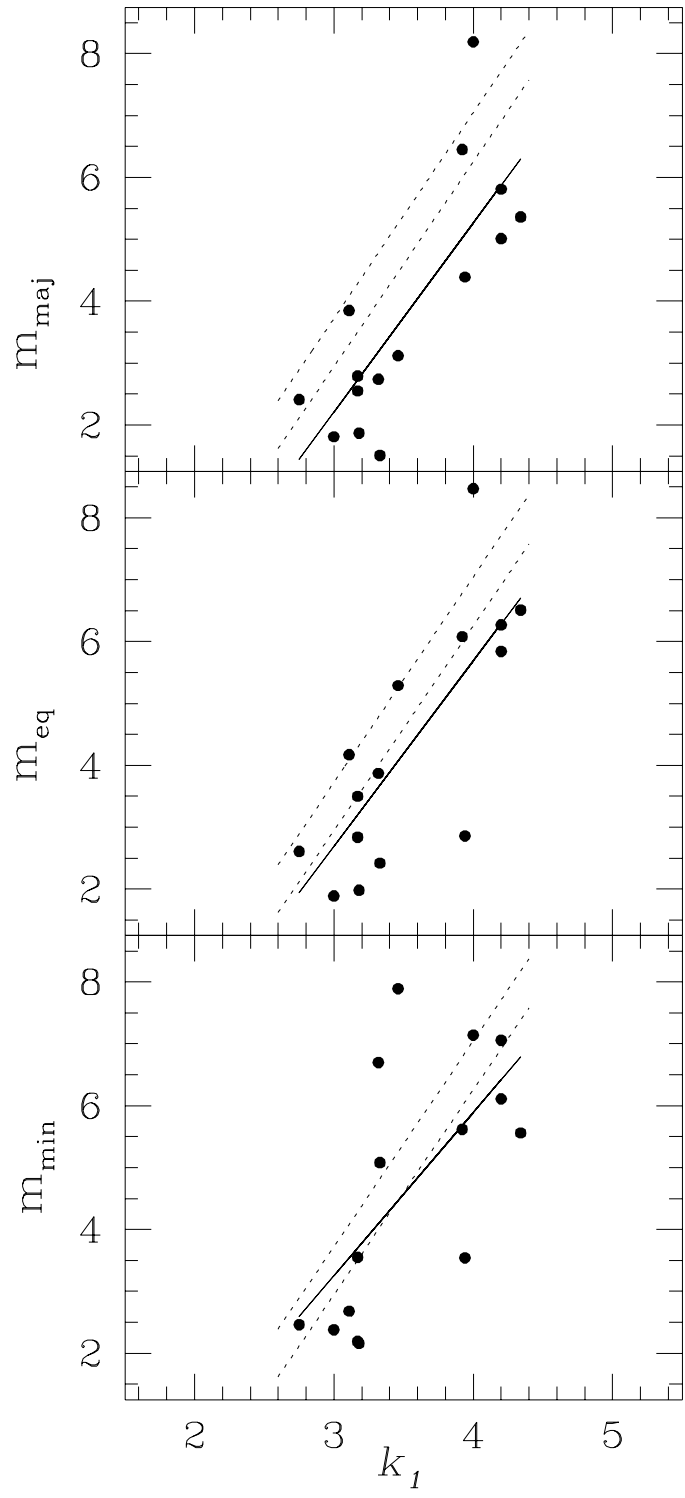
\bye